\documentclass[prd,nofootinbib,twocolumn,superscriptaddress]{revtex4}
\pdfoutput=1
\usepackage[T1]{fontenc}
\usepackage{amsmath,amssymb}
\usepackage{braket}
\usepackage{overpic}
\usepackage{graphicx}
\usepackage[usenames,dvipsnames]{color}
\usepackage{subfigure}
\usepackage{slashed}
\usepackage[colorlinks,citecolor=blue]{hyperref}
\usepackage{pdfpages}
\usepackage{color}
\usepackage{comment}
\usepackage{orcidlink}

\begin{document}



\title{LZ nuclear recoil event from inelastic singlet-doublet scalar dark matter}

\author{Disha Bandyopadhyay\orcidlink{https://orcid.org/0009-0002-0871-8538}}
\email{b.disha@iitg.ac.in}
\affiliation{Department of Physics, Indian Institute of Technology Guwahati, Assam 781039, India}

\author{Debasish Borah\orcidlink{https://orcid.org/0000-0001-8375-282X}}
\email{dborah@iitg.ac.in}
\affiliation{Department of Physics, Indian Institute of Technology Guwahati, Assam 781039, India}

\author{Pankaj Borah\orcidlink{https://orcid.org/0000-0003-2715-271X}}
\email{pankajborah@iitg.ac.in}
\affiliation{Department of Physics, Indian Institute of Technology
Guwahati, Assam 781039, India}

\begin{abstract}
We study the possibility of explaining the recently reported high-energy nuclear recoil event by the LUX-ZEPLIN (LZ) collaboration within the framework of singlet-doublet scalar dark matter (DM). Considering DM to be the CP even mass eigenstate formed out of a scalar doublet and a real scalar singlet, both being odd under an unbroken $Z_2$ symmetry, we find the parameter space of the model consistent with correct relic abundance and direct-detection limits on elastic scattering rates. A large part of the parameter space can also lead to inelastic up-scattering of DM with a rate and recoil energy consistent with the recent LZ event. Depending upon the singlet-doublet mixing, the model allows a much wider range of currently allowed parameter space compared to the pure scalar doublet DM limit, which is disfavored due to non-observations of high-energy neutrinos from solar-captured DM annihilation by the IceCube experiment. Presence of $Z_2$-odd right-handed neutrinos also leads to other interesting phenomenology related to the origin of light neutrino masses and leptogenesis.
\end{abstract}
\maketitle
\section{Introduction}
The LUX-ZEPLIN (LZ) experiment has reported a single high-energy nuclear recoil event using 2.84 ton-yr of exposure. The event corresponds to nuclear recoil energy of $248 \pm 23 (\rm stat) \pm 23 (\rm sys)$ keV which is in tension with the background-only hypothesis at a global significance of $2.6\sigma$ \cite{LZ:2026axp}. The event can be described by weakly interacting massive particle (WIMP) dark matter (DM) with masses above 200 GeV undergoing inelastic up-scattering off nucleons. Several DM interpretations\footnote{See \cite{Jeesun:2026vzo} for a non-DM explanation.} of the event have already appeared in the literature including Higgsino \cite{Fan:2026kxx, Wu:2026nhi, Freese:2026sga, Du:2026guj, Yin:2026jnn} as well as other DM candidates \cite{McCabe:2026crm, Unwin:2026rdp, Smirnov:2026aqk, Nomura:2026qyq, Lou:2026idn, Su:2026rwz, DiMauro:2026ldr, Yamashita:2026ump, Chattopadhyay:2026ryw, deLima:2026shq, Visinelli:2026kgt, Yin:2026jnn, Wang:2026ytg} or DM, in general \cite{Dent:2026bji, Gu:2026vto}. While the minimal higgsino DM may be tightly constrained or even disfavored due to solar-capture bounds \cite{Pospelov:2026ewn, Bose:2026ndd, Nguyen:2026lui, DiMauro:2026dqp} and non-observations of higher energy recoil events \cite{Rodd:2026tyn}, there exist non-minimal DM prescriptions which allow a larger allowed parameter space for DM mass and couplings. As the solar-capture bounds on such inelastic DM arise due to non-observations of high-energy neutrinos produced from DM annihilations into electroweak gauge boson pairs inside the Sun, it is motivating to consider DM with a singlet admixture instead of a pure electroweak doublet. This retains the inelastic nature of DM while weakening the solar-capture bounds and widening the parameter space consistent with relic-density and direct search limits.

In this work, we study the possibility of explaining this LZ event within the inelastic singlet-doublet scalar DM framework. The standard model (SM) is extended by a pair of $Z_2$-odd scalar doublet and real scalar singlet such that the lightest neutral eigenstate becomes the DM candidate. 
While the inert scalar doublet also offers the possibility of inelastic DM, fitting the correct relic and direct-detection bounds on elastic spin-independent (SI) scattering rates keep the parameter space tightly constrained \cite{Nomura:2026qyq, Wang:2026ytg}. Considering a singlet admixture offers more freedom in fitting the LZ data together with other constraints for a much wider parameter space in terms of DM mass and couplings. We calculate the LZ event rate for different DM mass and its mass splitting with the next-to-lightest $Z_2$-odd particle and the singlet-doublet mixing. In the appropriate limit of the singlet-doublet mixing, we recover the results for a scenario where DM is approximately a pure doublet. We also calculate the solar capture rate of DM and the resulting flux of neutrinos from DM annihilation inside the Sun. We then impose the limits from non-observation of high-energy neutrinos from the Sun by the IceCube experiment to find an upper bound on the singlet-doublet mixing, disfavoring the pure scalar doublet DM origin of the LZ event. We also briefly comment on the possibility of future probes related to DM as well as the connection of this setup to other unknowns like the origin of neutrino mass and baryon asymmetry of the Universe.

This paper is organized as follows. In section \ref{sec1}, we briefly discuss our model followed by the details of DM in section \ref{sec2}. In section \ref{sec3}, we discuss the details of inelastic DM scattering in the context of the recent LZ event followed by the details of indirect constraints from solar capture in section \ref{sec5}. We add our concluding remarks in section \ref{sec4}.


\section{The Model}
\label{sec1}
As mentioned earlier, we extend the SM by a real
scalar singlet $\varphi$ and an $SU(2)_L$ scalar doublet $\Phi$ with their quantum numbers being shown in Table \ref{tab:ScSM_fields}. All the new fields are odd under
an unbroken $Z_2$ symmetry, while the SM fields are
$Z_2$-even.
\begin{table}[h]
\centering
\begin{tabular}{c|cccc}
Field & $SU(3)_c$ & $SU(2)_L$ & $U(1)_Y$ & $Z_2$ \\ \hline
$\varphi$ & $\mathbf{1}$ & $\mathbf{1}$ & $0$ & $-$ \\
$\Phi$    & $\mathbf{1}$ & $\mathbf{2}$ & $1/2$ & $-$ \\
\end{tabular}
\caption{New particle content of the model.}
\label{tab:ScSM_fields}
\end{table}
The most general renormalizable scalar potential consistent with the SM gauge and
$Z_2$ symmetries contains the terms \cite{Beniwal:2020hjc}
\begin{align}
V \supset{}&
-\mu_H^2 H^\dagger H
+\lambda_H(H^\dagger H)^2
+m_\Phi^2\Phi^\dagger\Phi
+\lambda_\Phi(\Phi^\dagger\Phi)^2
\nonumber\\
&+\frac{1}{2}m_\varphi^2\varphi^2
+\frac{1}{4}\lambda_\varphi\varphi^4
+\lambda_{H\Phi,1}(H^\dagger H)(\Phi^\dagger\Phi)
\nonumber\\
&+\lambda_{H\Phi,2}(H^\dagger\Phi)(\Phi^\dagger H)
+\frac{1}{2}
\left[
\lambda_{H\Phi,3}(H^\dagger\Phi)^2
+\mathrm{H.c.}
\right]\nonumber\\
&
+\frac{1}{2}\lambda_{H\varphi}(H^\dagger H)\varphi^2
+\frac{1}{2}\lambda_{\Phi\varphi}
(\Phi^\dagger\Phi)\varphi^2
\nonumber\\
&+\left[
\kappa\,\Phi^\dagger H\,\varphi
+\mathrm{H.c.}
\right].
\label{eq:ScSM_potential}
\end{align}
where $H$ denotes the SM Higgs doublet and $\lambda_{H\Phi,3}$ controls the mass splitting between the CP-even and CP-odd components of $\Phi$, while $\kappa$ induces mixing between the CP-even component of the doublet and singlet scalars. After electroweak symmetry breaking, we write\\
\begin{equation}
\Phi =
\begin{pmatrix}
\phi^+\\
(\phi_R+iA)/\sqrt{2}
\end{pmatrix},
\end{equation}
\\
where $A$ is the CP-odd scalar. The CP-even states $\phi_R$ and
$\varphi$ mix according to
\begin{equation}
\label{eq:model:mixing}
\begin{pmatrix}
\eta_1\\
\eta_2
\end{pmatrix}
=
\begin{pmatrix}
\cos\theta & \sin\theta\\
-\sin\theta & \cos\theta
\end{pmatrix}
\begin{pmatrix}
\phi_R\\
\varphi
\end{pmatrix}.
\end{equation}
Without any loss of generality, we consider the lightest CP-even mass eigenstate $\eta_2$ to be our DM candidate. 

\section{Singlet-Doublet Scalar DM}
\label{sec2}
The CP-even scalar $\eta_2$, the DM candidate in our setup, is a mixture of the neutral CP-even component of the inert scalar doublet and the real scalar singlet, as defined in Eq.~(\ref{eq:model:mixing}). Thus, $\sin^2\theta$ measures the doublet fraction of the DM state: $\sin^2\theta = 1$ corresponds to the usual inert doublet limit \cite{Ma:2006km, Cirelli:2005uq, Barbieri:2006dq, LopezHonorez:2006gr,  Hambye:2009pw, Dolle:2009fn, LopezHonorez:2010eeh, LopezHonorez:2010tb, Borah:2012pu, Dasgupta:2014hha, Borah:2017dfn}, whereas $\sin^2\theta = 0$ gives a pure singlet DM \cite{Silveira:1985rk, McDonald:1993ex}. The mixing originates from the trilinear interaction $\kappa\,\Phi^\dagger H\varphi$, as already mentioned. The relic abundance of $\eta_2$ is determined by a combination of annihilation and coannihilation processes involving the $Z_2$-odd scalar sector. In particular, annihilation into electroweak gauge bosons, $\eta_2\eta_2\rightarrow W^+W^-,\,ZZ,$
is important whenever $\eta_2$ has an appreciable doublet component, while Higgs-mediated channels such as $\eta_2\eta_2\rightarrow hh,\; f\bar{f}$ can become relevant for sizeable scalar couplings. Coannihilations with the pseudoscalar $A$, the charged scalar $\phi^\pm$, and, when sufficiently close in mass, the heavier CP-even state $\eta_1$, can further affect the effective annihilation rate.

Note that, unlike in pure doublet DM, the annihilation strength of $\eta_2$ is no longer fixed entirely by its electroweak quantum numbers. The singlet admixture reduces the gauge interactions of the DM state, while the additional scalar interactions associated with $\varphi$ provide independent contributions to annihilation. Consequently, the relic-density constraint can be satisfied over a substantially broader region of parameter space than in the pure inert-doublet limit. In the latter case, gauge-dominated annihilation is sufficiently efficient that a scalar doublet saturating the observed abundance is typically pushed to masses above roughly\footnote{Although the singlet-doublet scalar DM admits relic-compatible regions below $550~\mathrm{GeV}$, we focus on $m_{\eta_2}\gtrsim 400~\mathrm{GeV}$ in the following, as this is the phenomenologically relevant high-mass regime for interpreting the $248~\mathrm{keV}$ LZ recoil through endothermic scattering.} $550~\mathrm{GeV}$. The singlet extension therefore provides additional freedom to accommodate the DM abundance without requiring the DM mass and electroweak interactions to remain tied to the pure-doublet prediction. We use \texttt{micrOMEGAs} \cite{Belanger:2026asz,Alguero:2023zol,Belanger:2018ccd} to compute the relic density of dark matter by incorporating all the relevant annihilation and coannihilation channels mentioned above.

Turning to direct detection (DD), the singlet-doublet mixing allows the elastic and inelastic scattering channels to be controlled differently. The $Z$-boson connects a CP-even to a CP-odd state of the same doublet, so there is no $Z\eta_2\eta_2$ vertex and elastic $Z$ exchange is absent. Conversely, $h$ is CP-diagonal, so there is no $h\eta_2 A$ vertex. The elastic SI rate is therefore purely Higgs-mediated, governed by the effective coupling
\begin{equation}
\lambda_{\rm eff} = \lambda_{H\varphi}\cos^2\theta -\frac{2\kappa\sin\theta\cos\theta}{v} +\lambda_{123}\sin^2\theta,
\end{equation}
with $\lambda_{123}\equiv \lambda_{H\Phi,1}+\lambda_{H\Phi,2}+\lambda_{H\Phi,3}$, giving $\sigma_{\rm SI}^{p} = \frac{\lambda_{\rm eff}^2 f_N^2\mu_p^2m_p^2}{4\pi m_h^4m_{\eta_2}^2}$.

The trilinear term with $\kappa$ enters with the opposite sign to the quartics, so $\lambda_{\rm eff}$ can be very small--a Higgs blind spot--suppressing $\xi\sigma_{\rm SI}^p$ below the current LZ limit and potentially even into the neutrino fog while leaving the inelastic channel unaffected. This cancellation is a distinctive feature of the singlet-doublet scalar DM and is absent for a pure doublet. Here $\xi\equiv\Omega_{\rm DM}h^2/\Omega^{\rm Planck}_{\rm DM}h^2$ accounts for a subdominant DM abundance.

\begin{figure}
    \centering
    \includegraphics[width=0.9\linewidth]{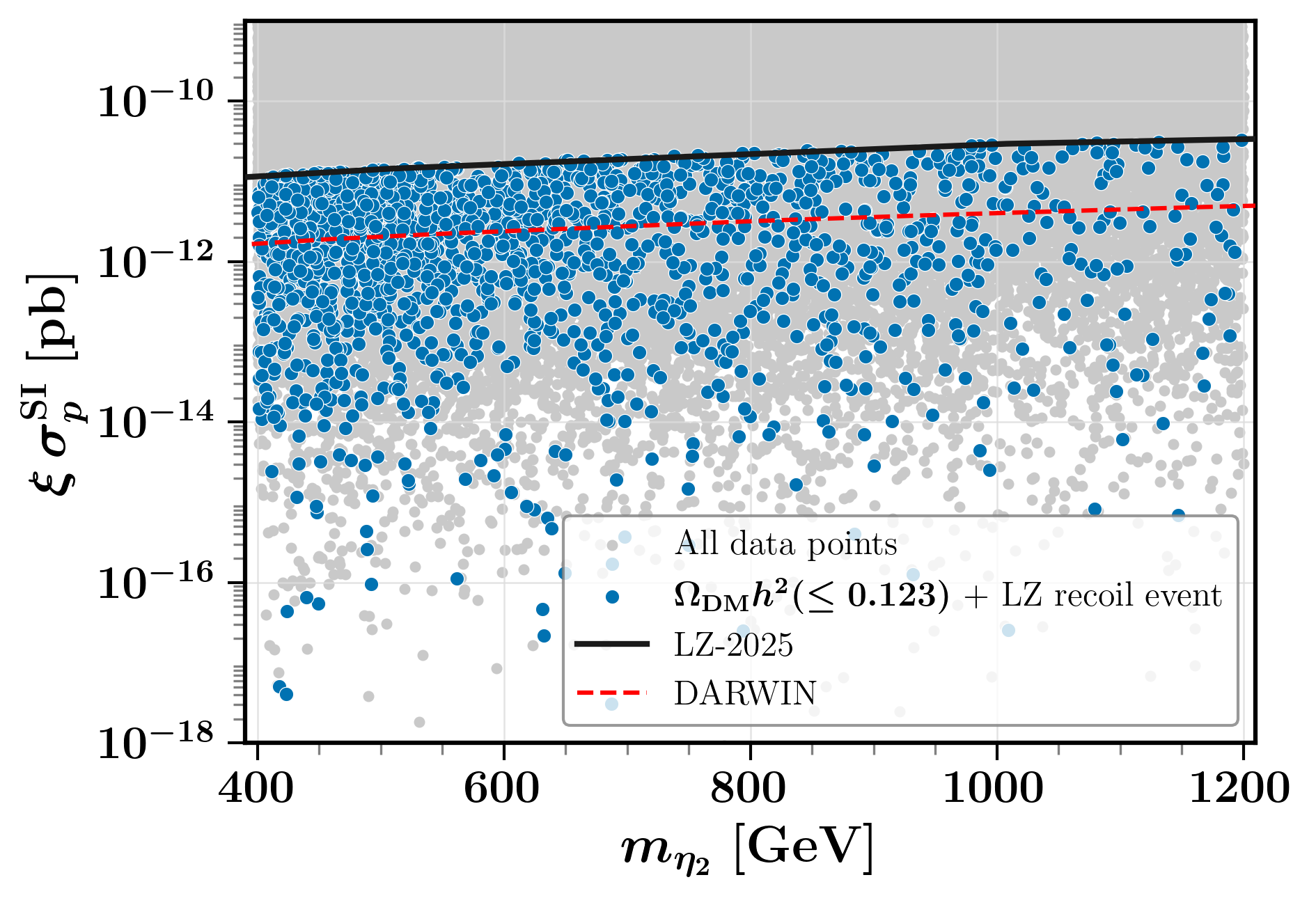}
    \caption{Spin-independent dark matter--proton scattering cross-section $\xi\,\sigma_p^{\rm SI}$ as a function of the dark matter mass $m_{\eta_2}$. Grey points show all scanned parameter points, while blue points satisfy the relic-density constraint within $3\sigma~(0.123)$ \cite{Planck:2018vyg} and the LZ recoil-event requirement $N_{\rm SR}=1$. The solid black curve denotes the LZ-2024 exclusion limit \cite{LZ:2026axp}.\label{fig:1}}
\end{figure}

We demonstrate the SI $\eta_2$–proton cross-section in Fig.~\ref{fig:1}, weighted by the relic fraction $\xi$, over the scan $m_{\eta_2}\in[400,1200]\text{ GeV}$, $\delta_1\equiv m_{\eta_1}-m_{\eta_2}\in[10,100]\text{ GeV}$, $\theta\in[0,\pi/2]$, and all independent quartic couplings in $[10^{-3},1]$. The model was implemented in \texttt{FeynRules} \cite{Alloul:2013bka}, relic abundance and direct-detection cross-sections were computed with \texttt{micrOMEGAs} \cite{Belanger:2026asz,Alguero:2023zol,Belanger:2018ccd}, and Higgs-sector constraints imposed via \texttt{HiggsBounds} \cite{Bechtle:2020pkv} and \texttt{HiggsSignals} \cite{Bechtle:2020uwn}. Blue points satisfy the relic-density bound within $+3\sigma$ ($\Omega h^2 = 0.120 \pm 0.001$) together with the LZ recoil requirement $N_{\rm SR} \simeq 1$. The solid black coloured curve is the LZ-2025 SI exclusion \cite{LZ:2024zvo} and the red coloured dashed curve is the projected DARWIN \cite{DARWIN:2016hyl} sensitivity. A substantial population survives well both the SI DD limits, spanning some eight decades in $\xi \sigma^{\rm SI}_p$ down to $\sim 10^{-8}~\text{pb}$. Since the trilinear $\kappa$ term enters $\lambda_{\rm eff}$ with sign opposite to the quartics and can drive it to a Higgs blind spot \cite{Beniwal:2020hjc}, suppressing the elastic channel while leaving the $Z$-mediated inelastic rate relevant to the $248~\text{keV}$ event untouched.

The inelastic channel is the one relevant to LZ. Only the doublet component of $\eta_2$ carries electroweak charge, so from the doublet kinetic term the neutral-current interaction is
\begin{equation}
\mathcal{L}_{Z}\supset -\frac{g}{2c_W}Z_\mu \left[ \cos\theta\, \bigl(\eta_1\overleftrightarrow{\partial^\mu}A\bigr) -\sin\theta\, \bigl(\eta_2\overleftrightarrow{\partial^\mu}A\bigr) \right],
\end{equation}
with $X\overleftrightarrow{\partial^\mu}Y \equiv X\partial^\mu Y-Y\partial^\mu X$. The $Z$ interaction is thus necessarily inelastic, $\eta_2+N\rightarrow A+N$, with $A$ the CP-odd partner, and its cross-section is controlled by the doublet fraction, $\sigma^{\rm inel.}_n\propto\sin^2\theta$. Thus, compared with pure doublet DM, the singlet-doublet scalar DM retains the weak-strength inelastic interaction while providing an additional handle on its normalization through the singlet admixture. The mixing angle therefore offers flexibility in accommodating the LZ rate for a given mass splitting $\delta_A\equiv m_A-m_{\eta_2}$, while the relic abundance remains subject to the full set of annihilation and coannihilation processes. Since $\delta_A/T_{\rm f.o.}\sim10^{-5}$, the $\mathrm{keV}$-scale splitting is irrelevant at freeze-out but decisive for terrestrial scattering: the small $\delta_A$ renders the $Z$-mediated process endothermic and confines it to the high-recoil region, precisely as required to interpret the $248~\mathrm{keV}$ event. The recoil kinematics and rate are developed in the next section.

\begin{figure*}[t]
    \centering
    \includegraphics[height=5.4cm,width=8.5cm]{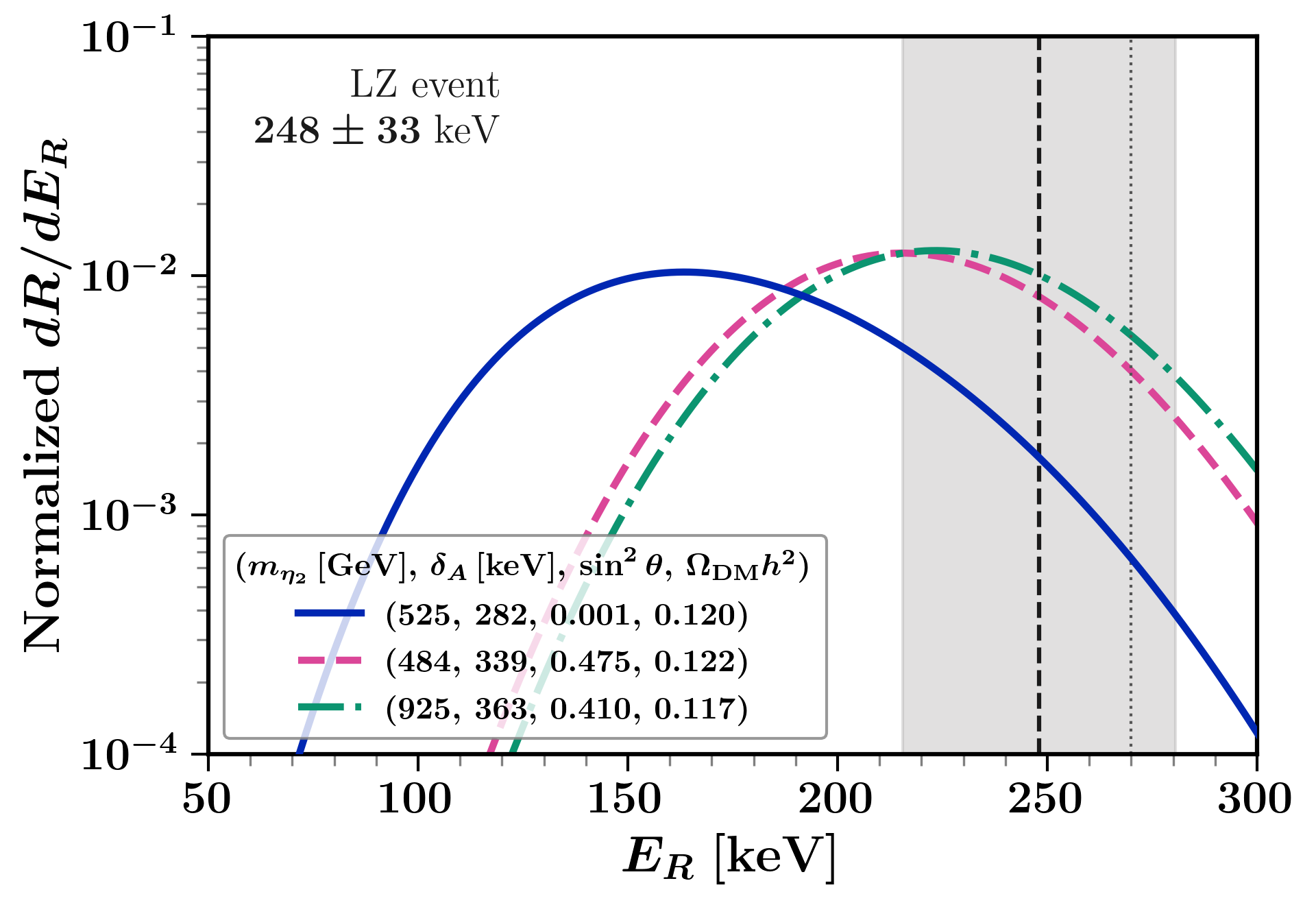}
    \hspace{0.2cm}
    \includegraphics[height=5.4cm,width=7.3cm]{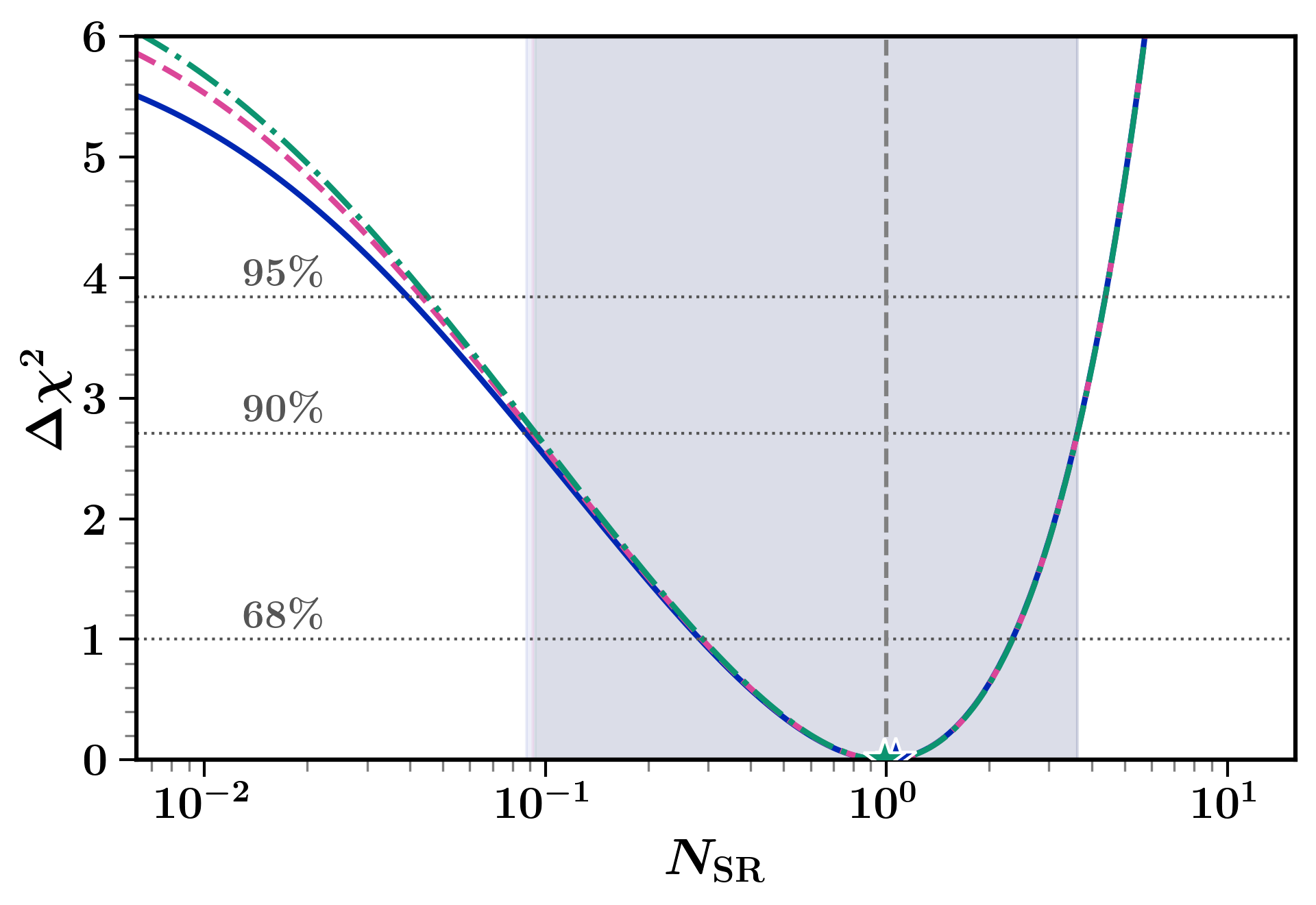}

    \includegraphics[height=5.4cm,width=8.5cm]{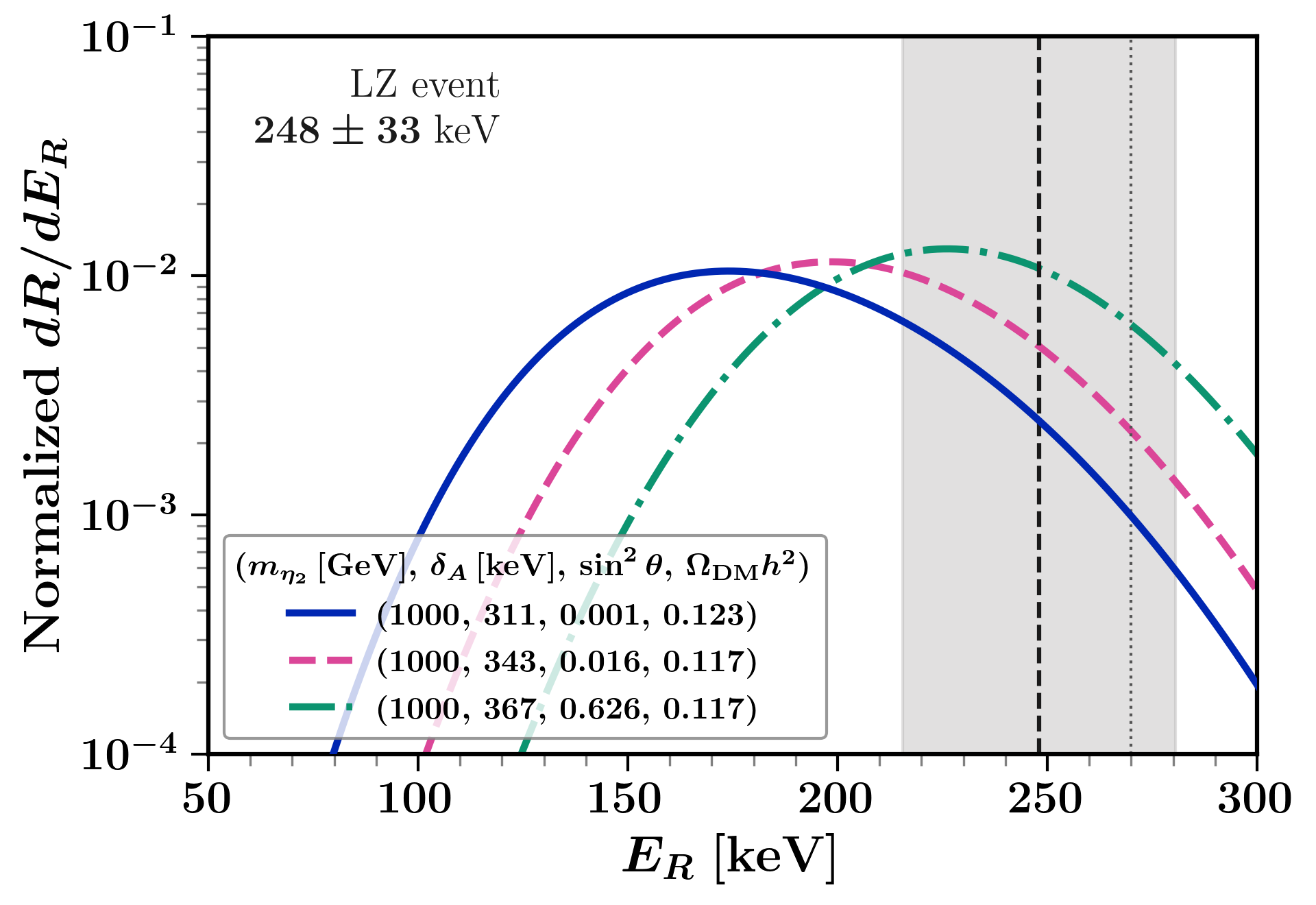}
    \hspace{0.2cm}
    \includegraphics[height=5.4cm,width=7.3cm]{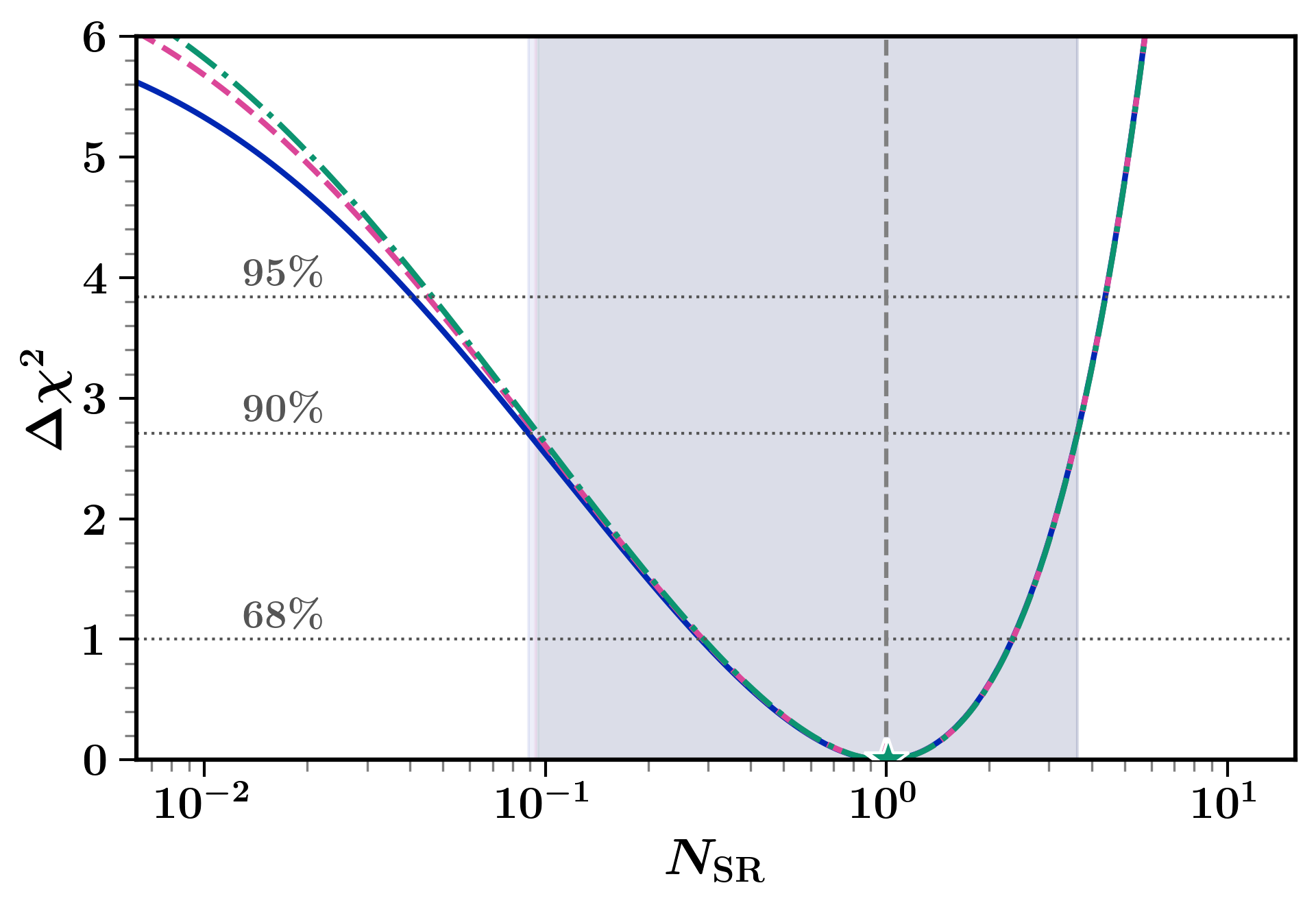}

    \vspace{-0.2cm}
    \makebox[0.48\textwidth][c]{(a)}
    \hfill
    \makebox[0.48\textwidth][c]{(b)}

    \caption{(a) Normalized differential recoil spectra, $dR/dE_R$, for six representative benchmark points in the inelastic dark matter parameter space. The vertical dashed line denotes the central recoil energy, $E_R=248~\mathrm{keV}$, of the LZ event, while the grey shaded band indicates the corresponds to the interval $E_R=248\pm33~\mathrm{keV}$ \cite{LZ:2026axp}. (b) $\Delta\chi^2$ as a function of the expected number of signal events, $N_{\rm SR}$, in the high-recoil signal region for the same benchmark points shown in panel (a). The shaded region denotes the range $0.1\lesssim N_{\rm SR}\lesssim3.6$ allowed at admitted at 90\% CL.}
    \label{fig:2}
\end{figure*}
%

\section{Origin of the LZ event}
\label{sec3}

%
In inelastic dark matter scattering, only dark matter particles with sufficiently large incident velocities can generate a recoil energy $E_R$. This recoil energy in the detector rest frame is given by
\begin{equation}
E_R(v,\theta)
=
\frac{2\mu_{\eta_2 X_i}^2v^2}{m_{X_i}}
\left[
1-\frac{v_t^2}{2v^2}
-\sqrt{1-\frac{v_t^2}{v^2}}\cos\theta'
\right],
\label{eq:ER_theta}
\end{equation}
where $v_t=\sqrt{\frac{2\delta_A}{\mu_{\eta_2 X_i}}}$ is the threshold velocity for the inelastic scattering, $\mu_{\eta_2 X_i}$ is the reduced mass of the $\eta_2$--$X_i$ system, $\mu_{\eta_2 X_i} = \frac{m_{\eta_2}m_{X_i}}{m_{\eta_2}+m_{X_i}}$ and $\theta'$
is the scattering angle in the center-of-mass frame. In the elastic limit $\delta_A\to 0$ (so $v_t\to 0$), this reduces to the standard result $E_R = (\mu_{\eta_2X_i}^2v^2/m_{X_i})(1-\cos\theta')$, with maximum recoil $E_R^{\max}=2\mu_{\eta_2X_i}^2v^2/m_{X_i}$ at $\cos\theta'=-1$. For a given incident velocity $v$, the recoil energy therefore lies in the range
$E_R^-(v)\leq E_R\leq E_R^+(v)$, corresponding to the two kinematic
limits $\cos\theta'=\pm1$.
For scattering off the $i$-th xenon isotope, denoted by $X_i$, the corresponding minimum velocity is \cite{Smith_2001, Tucker_Smith_2005}
\begin{equation}
v_{\min}^{(i)}(E_R)
=
\frac{1}{\sqrt{2m_{X_i}E_R}}
\left(
\frac{m_{X_i}E_R}{\mu_{\eta_2 X_i}}
+\delta_A
\right),
\label{eq:vmin_inelastic}
\end{equation}
Here, $X_i$ denotes the $i$-th xenon isotope and $m_{X_i}$ its corresponding nuclear mass. Unlike the elastic case, $v_{\min}^{(i)}(E_R)$ has an interior minimum, $v_{\min}^\star=v_t$, occurring at $E_R^\star=(\mu_{\eta_2X_i}/m_{X_i})\delta_A$, so the differential rate peaks near $E_R^\star$ rather than falling monotonically from threshold.\\

Note that, because the vector coupling of the $Z$ to nucleons is $g_V^n=-\frac{1}{2}$ and $g_V^p=\frac{1}{2}(1-4s_W^2)$, with $1-4s_W^2\simeq 0.075$, the proton contribution very nearly cancels: the interaction is neutron‑coupled rather than $A^2$‑coherent, and the coherent nuclear charge that appears is the weak charge $Q_W=(A-Z)-(1-4s_W^2)Z$, with $\langle Q_W^2\rangle/\langle A^2\rangle=0.3117$ for natural xenon. In the non‑relativistic limit, the zero‑momentum differential and total cross-sections on nucleus $X_i$ are $\frac{d\sigma_{X_i}}{dE_R}=\frac{G_F^2 m_{X_i} Q_W^2 \sin^2\theta}{4\pi} F^2(E_R)$ and $\sigma_{X_i}^0=\frac{G_F^2\mu_i^2 Q_W^2}{2\pi}\sin^2\theta$, such that the corresponding per‑neutron cross-section is
\begin{equation}
\label{eq:sigma-n}
    \sigma^{\rm inel.}_n = \frac{G_F^2\mu_n^2}{2\pi}\sin^2\theta = 7.4\times 10^{-39}\sin^2\theta\text{ cm}^2,
\end{equation}
rescaled by the doublet fraction $\sin^2\theta$.

The differential recoil rate in the Earth frame is given by
\begin{equation}
\frac{dR}{dE_R} = \frac{\rho_0 \xi}{m_{\eta_2}}\sum_i \frac{f_i}{m_{X_i}} \int_{v>v_{\min}^{(i)}} d^3v \, v \, f_E(\mathbf{v},t) \frac{d\sigma_{X_i}}{dE_R}(v,E_R),
\end{equation}
where $f_E(\mathbf{v},t)$ is the dark matter velocity distribution in the Earth frame, $\rho_0$ is the local dark matter energy density, and $f_i$ denotes the natural abundance of the $i$‑th xenon isotope with $\sum_i f_i=1$, and $\xi$ is defined in Sec.~\ref{sec2}. In the non-relativistic limit, the differential cross-section for inelastic dark matter--nucleus scattering can be written as
\begin{equation}
\frac{d\sigma_{X_i}}{dE_R}(v,E_R)
=
\frac{1}{32\pi}
\frac{1}{m_{\eta_2}^2m_{X_i}}
\frac{1}{v^2}
\left|\mathcal{M}_{X_i}\right|^2,
\label{eq:diff_cross_section}
\end{equation}
where $\mathcal{M}_{X_i}$ denotes the $\eta_2$--xenon-nucleus scattering matrix element, from which the zero‑momentum cross-section $\sigma_{X_i}^0$ above follows directly.

\begin{figure*}[!t]
    \centering
   \includegraphics[width=0.48\linewidth]{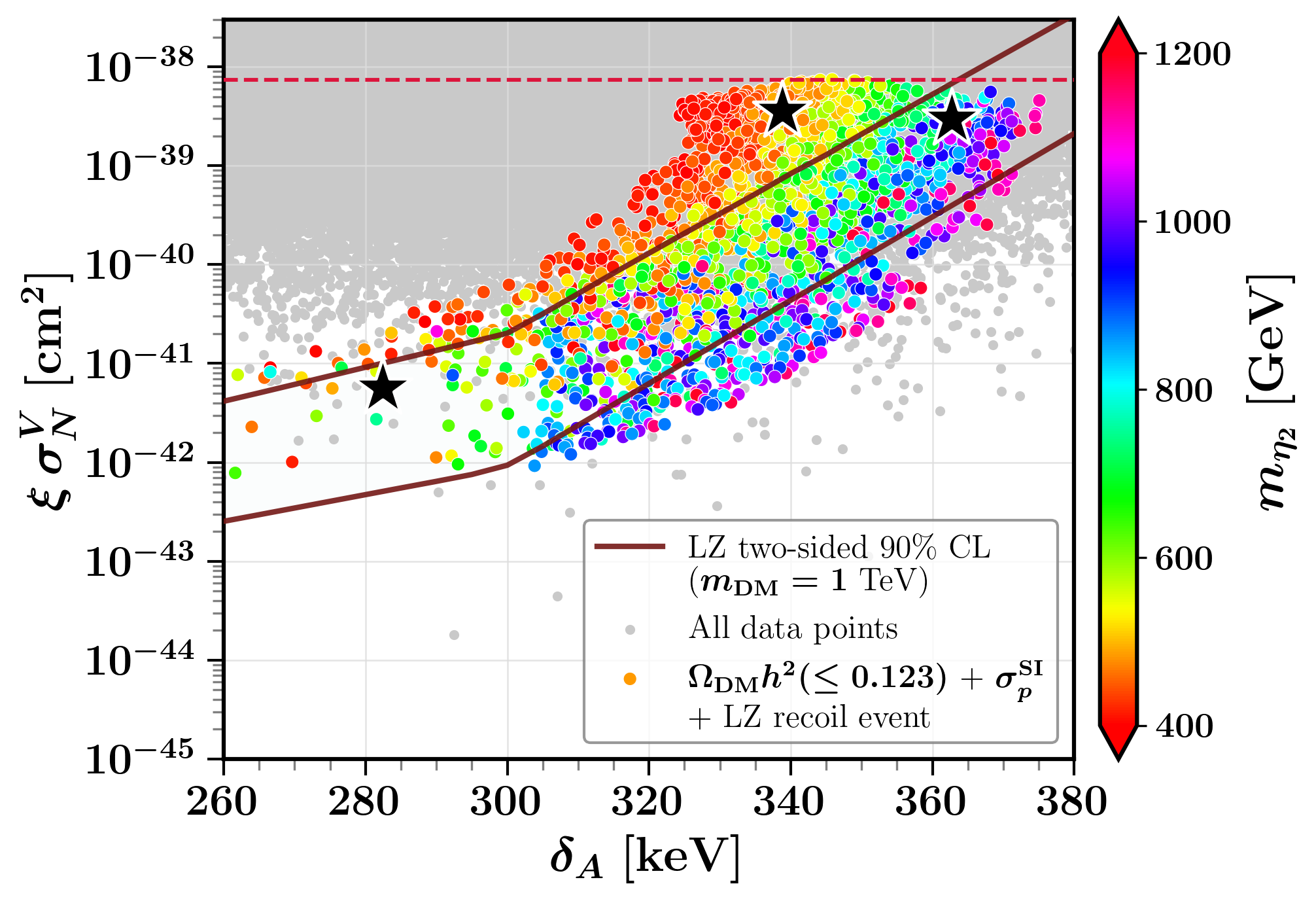}
    \hfill
    \includegraphics[width=0.48\textwidth]{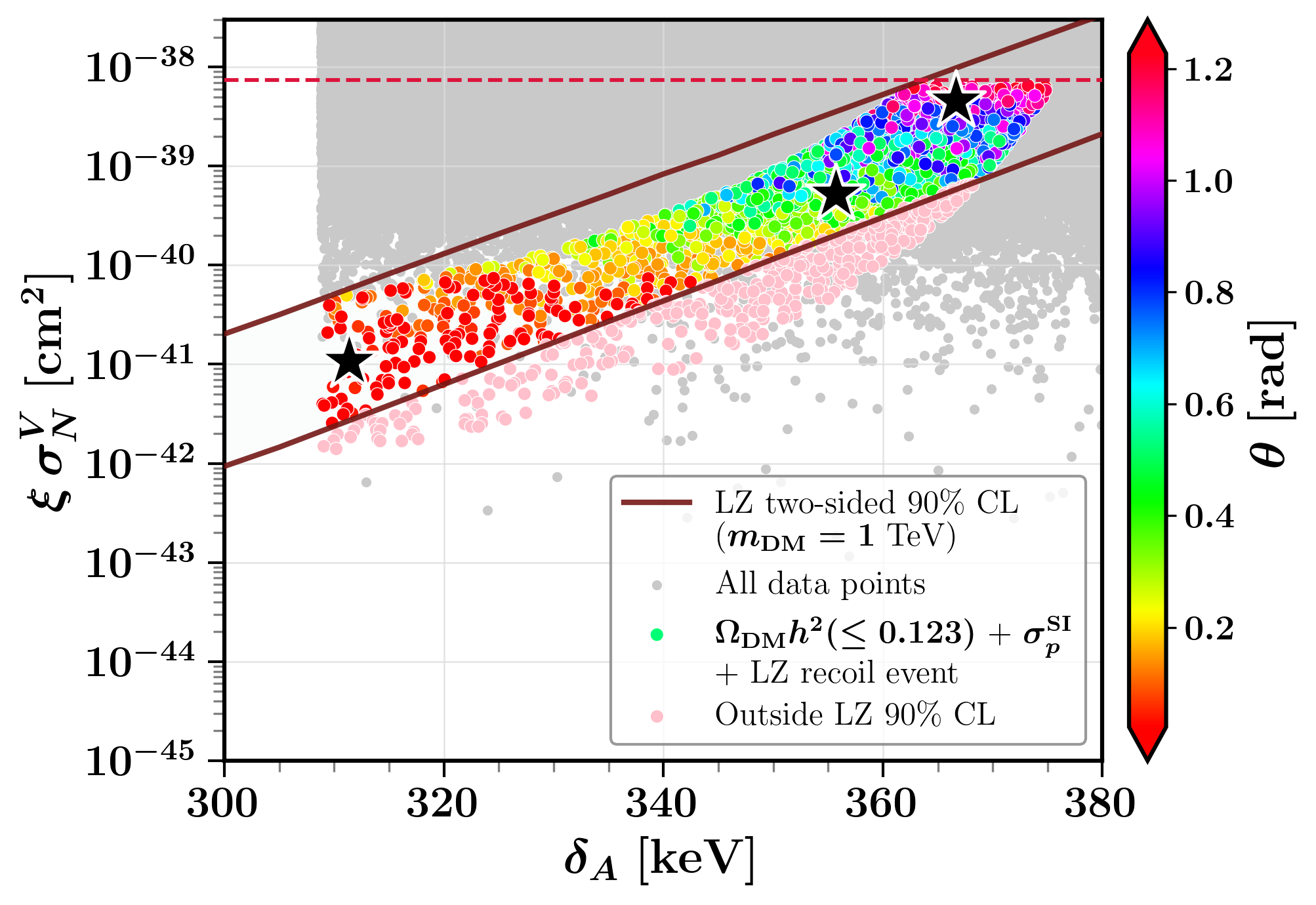}

    \vspace{-0.2cm}
    \makebox[0.48\textwidth][c]{(a)}
    \hfill
    \makebox[0.48\textwidth][c]{(b)}

    \caption{DM-fraction-weighted inelastic $\eta_2$--nucleon cross-section, $\xi\,\sigma_N^V$, as a function of the mass splitting $\delta_A$. The dark-red dashed line corresponds to the ceiling $\xi\sin^2\theta=1$, while grey points show the full scan and coloured points satisfy the relic-density, LZ SI limit on $\sigma_p^{\rm SI}$ and $N_{\rm SR}=1$ constraints. The dark-brown colored curves are the LZ two-sided 90\% CL interval at $m_{\rm DM}=1\text{ TeV}$, and the $\bigstar$-shaped points are the benchmarks of Fig.~\ref{fig:2}. In panel (a), $m_{\eta_2}$ is varied over $[400,1200]~\mathrm{GeV}$, with the colour indicating the DM mass, while in panel (b) $m_{\eta_2}=1~\mathrm{TeV}$ is fixed and the colour denotes the mixing angle $\theta$.}
    \label{fig:3}
\end{figure*}

We adopt the Standard Halo Model (SHM)  \cite{Baxter_2021,Smith:2006ym,LEWIN199687} with
\begin{equation}
\rho_0=0.3~{\rm GeV\,cm^{-3}}, 
\, \,
v_0=238~{\rm km\,s^{-1}},
\, \,
v_{\rm esc}=544~{\rm km\,s^{-1}}.
\end{equation}
At the rate level, the magnitude of the laboratory velocity relative to the Galactic frame is approximated as
\begin{equation}
v_E(t)
=
254~{\rm km\,s^{-1}}
+
15~{\rm km\,s^{-1}}
\cos[\omega(t-t_0)],
\label{eq:earth_velocity}
\end{equation}
where $t_0$ corresponds to the annual maximum in early June. The annual-averaged recoil spectrum is defined as $\left\langle\frac{dR}{dE_R}\right\rangle = \frac{1}{T} \int_0^T dt\, \frac{dR}{dE_R}(t).$ 

For an exposure of $\mathcal{E}=2.84\text{ tonne yr}$ \cite{LZ:2026axp}, the expected number of signal events in the extended LZ recoil-energy window is estimated as,
\begin{equation}
    N_{\rm SR} = \mathcal{E}\int_{5.4\text{ keV}}^{269.9\text{ keV}} dE_R\,\epsilon_{\rm LZ}(E_R)\left\langle\frac{dR}{dE_R}\right\rangle,
\end{equation}
where $\epsilon_{\rm LZ}(E_R)$ denotes the LZ detection efficiency, which is 50\% at $E_R=5.4\text{ keV}$, reaches a mean of 96\% over 14–$250\text{ keV}$, and falls back to 50\% at $E_R=269.9\text{ keV}$ \cite{LZ:2026axp}. For the corresponding to LZ's $500 < S1c < 600$ phd panel, the predicted signal yield is evaluated in the high‑energy signal region $200\text{ keV}\le E_R\le269.9\text{ keV}$, in which the expected background is $b=0.0106\pm0.0008$ counts \cite{LZ:2026axp}. We convolve the predicted rate with LZ’s region of interest (ROI) efficiency and with a Gaussian energy response of width $\sigma_E(E_R) = \left[ \frac{(23\text{ keV})^2 E_R}{248\text{ keV}} + (0.093 E_R)^2 \right]^{1/2}$. Finally, we assess compatibility with the LZ event through an extended unbinned likelihood for one observed event at $E_{\rm obs}=248\text{ keV}$ following a approach similar to \cite{Wu:2026nhi, Su:2026rwz},
\begin{equation}
  -2\ln\mathcal{L}(N_{\rm SR}) = 2(N_{\rm SR}+b) - 2\ln\bigl[N_{\rm SR}\,f_s(E_{\rm obs}) + b\,f_b\bigr],
\end{equation}
with $f_s(E)\equiv N_{\rm SR}^{-1}\,dN_{\rm obs}/dE$ the normalized signal shape and $f_b=1/(269.9-200)\text{ keV}^{-1}$, and we define $\Delta\chi^2\equiv -2\ln\mathcal{L}(N_{\rm SR})+2\ln\mathcal{L}(N_{\rm SR}^{\rm best})$, with $\Delta\chi^2 < 2.71$ the 90\% CL region for one degree of freedom. For any maximum‑likelihood point $N_{\rm SR}^{\rm best}\to 1.0$, the requirement $N_{\rm SR}=1$ is an accurate proxy for the best fit. For each value of the DM mass $m_{\eta_2}$ satisfying the relic‑abundance and SI DD constraint, the inelastic mass splitting $\delta_A=m_A-m_{\eta_2}$ is therefore chosen such that the expected number of signal events in the extended LZ recoil window is unity, i.e., $N_{\rm SR}=1$




Fig. \ref{fig:2} shows, in panels (a), the normalized differential recoil spectra $dR/dE_R$ for six representative benchmarks in two mass groups ($m_{\eta_2} \in [400,1200]~\text{GeV}$, top; fixed at $1\text{ TeV}$, bottom), each labelled by $(m_{\eta_2},\delta_A,\sin^2\theta,\Omega_{\rm DM}h^2)$. The predicted rate is weighted by the LZ nuclear-recoil efficiency $\epsilon_{\rm LZ}(E_R)$ and convolved with the detector energy response, a Gaussian of width $\sigma_E(E_R) = \left[(23\text{ keV})^2 \frac{E_R}{248\text{ keV}} + (0.093 E_R)^2\right]^{1/2}$, so that the abscissa is the reconstructed recoil energy directly comparable to the observed event; the curves are then normalized to unit integral over the analysis window $E_R\in[5.4, 269.9]\text{ keV}$. The vertical dashed line and grey band denote the LZ feature at $E_R=248\pm 33\text{ keV}$. Accurate prediction of the scattering rate on xenon depends sensitively on the nuclear form factor $F(q)$. The relevant momentum transfers for this signal lie in the regime $q \simeq 1.2-1.6\text{ fm}^{-1}$, probing the region around and beyond the second diffraction minimum. For our fiducial rate calculations, we adopt the Helm parameterization in its standard Lewin–Smith form \cite{Lewin:1995rx}\footnote{We also cross-checked and verified our results using publicly available nuclear response functions \cite{WIMpy-code, Jeong:2021bpl}.}.

In Fig. \ref{fig:2}(a), each spectrum rises through the region of interest, driven by the endothermic threshold kinematics that confine the signal to high recoil energy, and peaks near the observed event before falling. A large fraction of the in-window rate, $\sim 60-80\%$, lies within the signal region, consistent with a spectrum weighted toward high recoil energy as required to accommodate the $248\text{ keV}$ feature. The three benchmark spectra in each panel exhibit similar shapes and comparable rates across the signal region, although their peak positions differ appreciably, particularly for the navy-blue spectra. These differences reflect the dependence of the recoil spectrum on the DM mass and mass splitting, while all benchmarks retain sufficient spectral weight around $248\,\mathrm{keV}$ to accommodate the observed event.
The corresponding spectra without the Gaussian energy response applied are shown in Appendix~\ref{appx:true-recoil} (Fig.~\ref{fig:appx:true-recoil}); there the bare threshold kinematics, smoothed here by the finite detector resolution, become manifest.

Panels (b) of Fig. \ref{fig:2} show the profiled $\Delta\chi^2$ as a function of the expected signal count $N_{\rm SR}$ in the high-recoil signal region, obtained from the extended likelihood with the single observed event and its estimated background. The three horizontal lines mark the 68\%, 90\%, and 95\% confidence levels. The shaded band indicates the interval $0.1\lesssim N_{\rm SR}\lesssim 3.6$ admitted at 90\% CL, and the minimum near $N_{\rm SR}\simeq 1$ corresponds to a single expected count, as required to accommodate the observation. The $\Delta\chi^2$ curves for all benchmarks coincide, reflecting that with one event and negligible background the constraint acts on the overall normalization alone and is essentially independent of spectral shape.

In Fig.~\ref{fig:3}, we show the inelastic $\eta_2$–nucleon cross-section weighted by the DM fraction, $\xi\,\sigma_N^V$, against the mass splitting $\delta_A$. Here, $\sigma_N^V = \frac{\mu_n^2 (c_1^s m_v^2)^2}{\pi m_v^4}$ up to the isoscalar normalisation \cite{LZ:2026axp}, and equivalent to $\sigma^{\rm inel.}_n$, Eq.~(\ref{eq:sigma-n}). The horizontal dark red coloured dashed line marks the ceiling $\xi\sin^2\theta=1$, i.e. a pure inert doublet saturating the relic density; every point in the singlet-doublet model lies below it, with the vertical displacement measuring the singlet admixture through $\xi\sin^2\theta$. Grey points are the full scan, coloured points satisfy the relic bound within $0.123~(+3\sigma)$, the LZ SI limit on $\sigma_p^{\rm SI}$, and $N_{\rm SR}=1$. The dark-brown colored curves are the LZ two-sided 90\% CL interval at $m_{\rm DM}=1\text{ TeV}$, and the $\bigstar$-shaped points are the benchmarks of Fig.~\ref{fig:2}. In panel (a) the DM mass is scanned over $m_{\eta_2}\in[400,1200]\text{ GeV}$ (colour), giving $\delta_A\in[271.6,374.6]\text{ keV}$ and $E_R^\star\in[210,338]\text{ keV}$. Whereas, in panel (b) the mass is fixed at $1\text{ TeV}$ and the colour encodes the mixing angle $\theta$, narrowing the viable range to $\delta_A\in[309.8,374.8]\text{ keV}$ and $E_R^\star\in[276,333]\text{ keV}$. (Since $\mu_n$ saturates above $\sim 100\text{ GeV}$, the conversion above is mass-independent to better than 0.5\%; the LZ band itself, however, is quoted at $1\text{ TeV}$, so panel (a) should be read as indicative away from that mass.)

The two-sided nature of the LZ interval is what makes the singlet admixture essential. The band rises steeply with $\delta_A$—by roughly three decades between $310$ and $375\text{ keV}$—because a larger splitting suppresses the rate kinematically and must be compensated by a larger cross-section. A pure inert doublet, by contrast, has no freedom in the normalisation: once the relic density fixes the mass, $\xi\sin^2\theta=1$ is not a choice, and the prediction is the single horizontal line, which intersects the allowed band only in the narrow corner $\delta_A\gtrsim 360\text{ keV}$, uncomfortably close to the kinematic wall $\delta_A\to\delta_{\max}$ and correspondingly sensitive to $v_{\rm esc}$. In the singlet-doublet DM model the mixing angle decouples the normalisation from the kinematics. Thus, $\xi\sin^2\theta$ spans four decades $\sim 10^{-4} - 1$, and the surviving points populate the band continuously across the full range $\delta_A\simeq 310-375\text{ keV}$. Panel (b) makes the mechanism explicit--the colour ordering is monotonic in $\theta$, with singlet-dominated points ($\theta\lesssim 0.3$, $\sin^2\theta\lesssim 0.09$) falling near $10^{-40}\text{ cm}^2$ and doublet-dominated points ($\theta\gtrsim 0.9$) approaching the tree-level ceiling, while the pink points lie outside the 90\% CL. Consistently, the benchmarks closed to the dashed line is the most doublet-like ($\sin^2\theta=0.626$), and the lowest star the most singlet-like ($\sin^2\theta=0.001$). Notably, even the former retains a $\sim 40\%$ singlet component. The singlet therefore does two things at once: it relaxes the relic-density requirement, allowing $\eta_2$ to be a thermal candidate over a much wider mass range than the $\gtrsim 550\text{ GeV}$ demanded of a pure doublet, and it supplies the independent handle on $\sigma_N^V$ needed to place the predicted rate inside a two-sided measurement rather than merely below an upper limit—which is precisely what interpreting a positive recoil event, as opposed to setting an exclusion, requires.
\section{Indirect Constraints from Solar Capture and Annihilation}
\label{sec5}
\begin{figure}[!t]
    \centering
    \begin{overpic}[height=5.8cm, width=8.4cm]{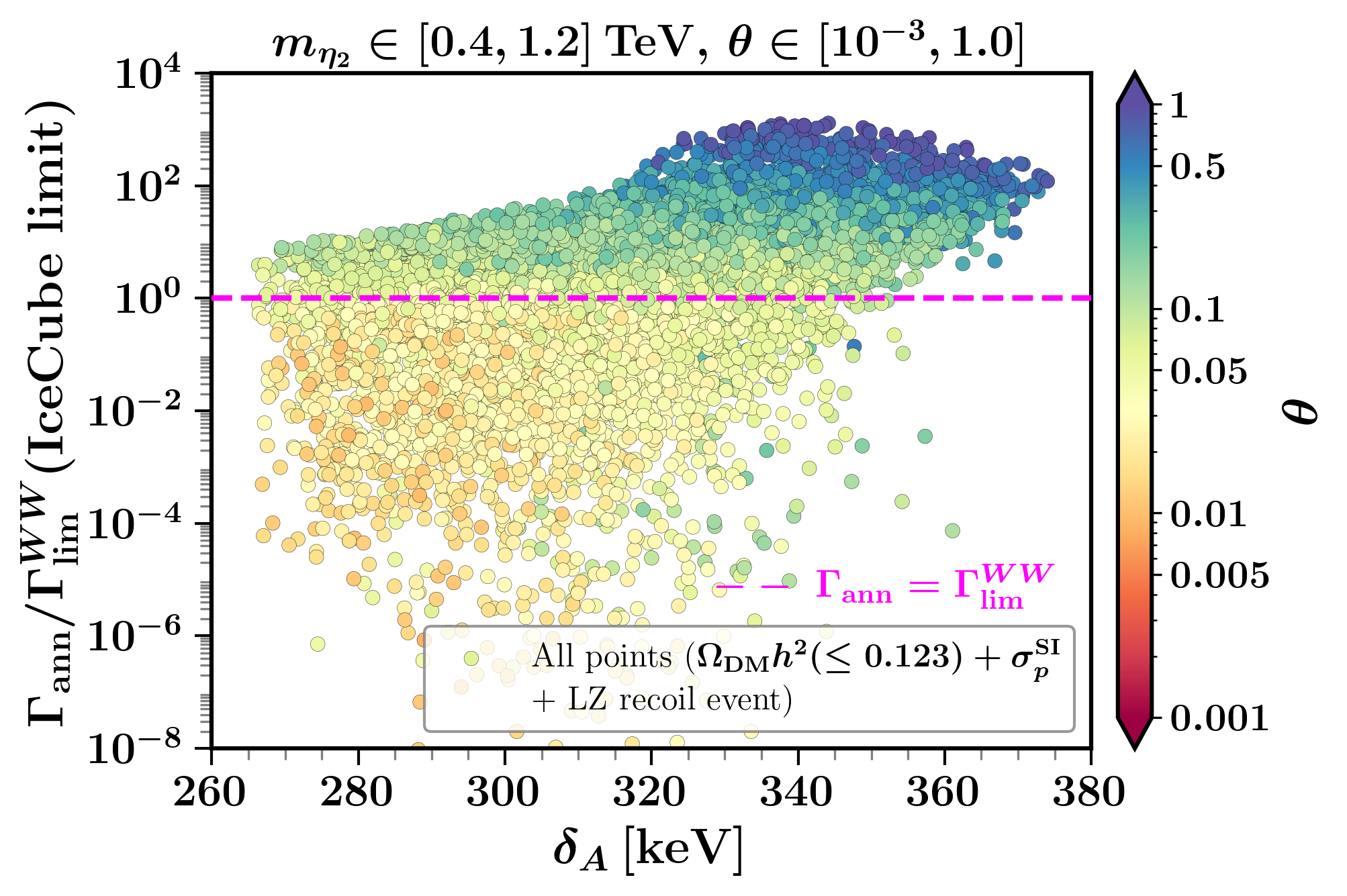}
    \end{overpic}
    \caption{Ratio of the solar annihilation rate to the adopted IceCube $W^+W^-$ limit, $\Gamma_{\mathrm{ann}}/\Gamma_{\mathrm{lim}}^{WW}$, as a function of the inelastic DM mass splitting $\delta_A$, with the color scale indicating the singlet-doublet mixing angle $\theta$. The dashed magenta line denotes $\Gamma_{\mathrm{ann}}/\Gamma_{\mathrm{lim}}^{WW} = 1$, above which the corresponding parameter points exceed the adopted IceCube limit. All points satisfy the LZ recoil-event normalization and correspond to $ \Omega_{\rm DM} h^2 \leq 0.123$.}
    \label{fig:solar-theta}
\end{figure}

\begin{figure*}[!htpb]
    \centering
   \includegraphics[width=0.48\linewidth]{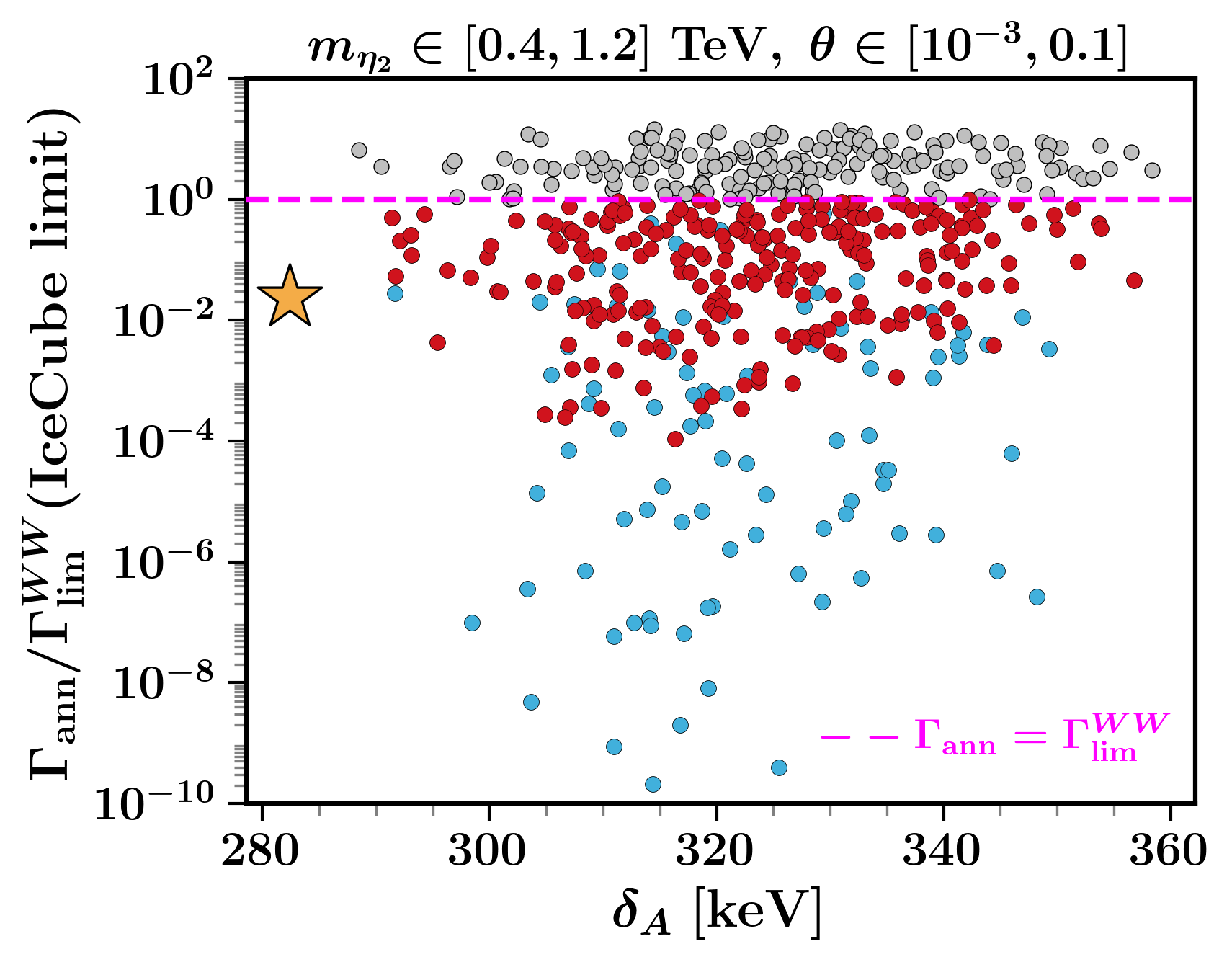}
    \hfill
    \includegraphics[width=0.48\textwidth]{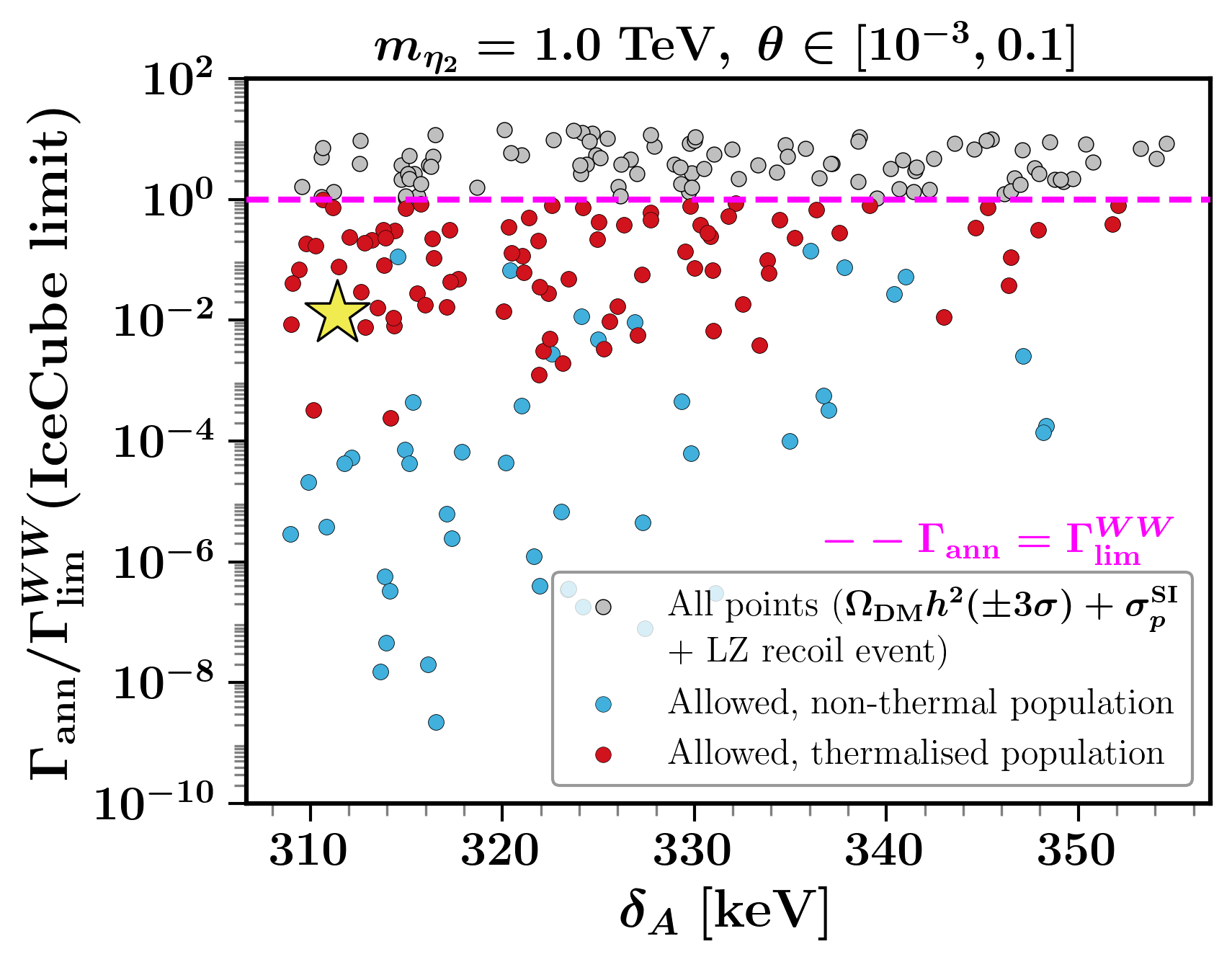}

    \vspace{-0.2cm}
    \makebox[0.48\textwidth][c]{(a)}
    \hfill
    \makebox[0.48\textwidth][c]{(b)}

    \caption{Ratio of the solar annihilation rate to the representative IceCube limit \cite{IceCube:2025fcu}, $\Gamma_{\mathrm{ann}}/\Gamma_{\mathrm{lim}}^{WW}$, as a function of the inelastic DM mass splitting $\delta_A$. All the points shown satisfy the $\pm 3\sigma$ relic-density constraint and the LZ recoil event. The grey points correspond to parameter points that exceed the adopted IceCube limit, while the blue and red points correspond to non-thermalized and fully thermalized captured DM populations, respectively. The horizontal dashed line denotes $\Gamma_{\mathrm{ann}} = \Gamma_{\mathrm{lim}}^{WW}$. Panel (a) and (b) correspond to the cases with varying DM mass and fixed $m_{\eta_2} = 1$ TeV, respectively. The two benchmark points discussed in the text are indicated by stars.}
    \label{fig:ID}
\end{figure*}
\begin{table*}[!htpb]
\centering
\begin{tabular}{c c c c c c c c c c}
\hline\hline
 & $m_{\eta_2}$ [GeV] & $\delta_A$ [keV] & $\sin^2\theta$ & $\Omega h^2$ & $\langle\sigma v\rangle$ [cm$^3$/s] & $C$ [s$^{-1}$] & $t_\odot/\tau$ & $\tanh^2(t_\odot/\tau)$ & $\Gamma_{\rm ann}/\Gamma_{\rm lim}^{WW}$ \\
\hline
BP-I  & 524.94 & 282.39 & $7.63\times10^{-4}$ & 0.1201 & $4.97\times10^{-31}$ & $4.132\times10^{20}$ & 0.0857 & $7.32\times10^{-3}$ & 0.0205 \\
BP-II & 1000.0 & 311.38 & $1.40\times10^{-3}$ & 0.1227 & $4.76\times10^{-31}$ & $1.748\times10^{20}$ & 0.0885 & $7.79\times10^{-3}$ & 0.0110 \\
\hline\hline
\end{tabular}
\caption{Benchmark points highlighted (star markers) in Fig.~\ref{fig:ID}(a) (orange colored) and~\ref{fig:ID}(b) (yellow colored), respectively. Both are fully thermalized captures ($V_{\rm eff}=V_{\rm eff}^{\rm th}$) and safely below the IceCube limit.}
\label{tab:benchmarks}
\end{table*}

The inelastic interaction responsible for the high-recoil LZ event can also lead to gravitational capture of DM in the Sun. In contrast to terrestrial scattering, the DM velocity inside the Sun is enhanced by gravitational acceleration, $w^2(r) = u^2 + v_{\mathrm{esc}}^2(r)$, where $u$ is the DM velocity far from the Sun. The capture rate is therefore obtained by integrating the inelastic scattering probability over the solar density profile and the halo velocity distribution, given by the Gould integral \cite{Press:1985ug,Gould:1987ir},
\begin{eqnarray}
    C_\odot &= \sum_A {\Large \int}_0^{R_\odot} 4\pi r^2 \, dr\, n_A(r) {\Large \int}_0^{u_{\mathrm{max}}} du \frac{d\Phi}{du} \nonumber \\ 
    &\times {\Large \int}_{E_R^{\mathrm{min}}}^{E_R^+} dE_R \frac{d\sigma_{N}}{dE_R},
\end{eqnarray}
where $\frac{d\Phi}{du} = \frac{\rho_0}{m_{\eta_2}} F(u) \frac{w^2(r,u)}{u}$, and $u_{\mathrm{max}}$ denotes the maximum asymptotic DM speed that allows capture at radius $r$. The function $F(u) \equiv u^2 \int d\Omega_u f_u(\vec{u})$ represents the angle-averaged speed distribution in the Sun's rest frame, with $f_u(\vec{u}) = f_{\mathrm{MB}}(\vec{u} + \vec{v}_\odot)$ accounting for the Sun's motion relative to the Galactic halo \cite{Pospelov:2026ewn}. The lower integration limit $E_R^{\mathrm{min}} = \max\left[ E_R^-, ~\frac{1}{2} m_{\eta_2} u^2 - \delta_A \right]$ is the minimum recoil energy required for capture. The differential scattering cross-section $d\sigma_N / dE_R$ is suppressed by the singlet-doublet mixing squared, $\sigma_N^{\mathrm{inel}} \propto \sin^2\theta$, where $N$ is the nuclear species. Although the endothermic energy threshold kinematically suppresses capture, the boosted interior solar velocity $w(r,u)$ enables capture on sufficiently heavy nuclei. We use a BS05-like \cite{Bahcall:2004pz} solar density/composition profile; for the LZ-preferred window capture is dominated by iron \cite{DiMauro:2026dqp, Lee:2026jxl}.

Once captured, the DM population evolves according to capture and annihilation. The annihilation rate can be written as
\begin{equation}
\Gamma_{\mathrm{ann}} = \frac{1}{2} C_{\odot} \tanh^2\left(\frac{t_{\odot}}{\tau}\right), \quad \frac{t_{\odot}}{\tau} = t_{\odot} \sqrt{\frac{C_{\odot} \langle \sigma v \rangle_{\mathrm{tot}}}{V_{\mathrm{eff}}}}.
\end{equation}
with $\langle\sigma v\rangle_{\rm tot}$ the total present-day annihilation cross-section, and the effective volume $V_{\mathrm{eff}}$ characterizes the spatial distribution of the captured population. Inelastic down-scattering on iron (Fe) stalls below $E_{\mathrm{stall}} = m_{\eta_2}\delta_A/\mu_{\mathrm{Fe}}$; further cooling proceeds only through the Higgs-mediated elastic channel governed by $\sigma_{\mathrm{SI}}^p$ and we evolve the resulting orbit to determine whether the population thermalizes ($V_{\mathrm{eff}} \to V_{\mathrm{eff}}^{\mathrm{th}}$) or remains extended, following the ansatz of \cite{Lee:2026jxl}.

Captured DM can annihilate into $W^+W^-, ~ZZ, ~hh$, etc., sourcing a high-energy solar neutrino flux constrained by IceCube \cite{IceCube:2025fcu}. Using the ten-year IceCube solar search, Ref. \cite{Pospelov:2026ewn} derived a representative
limit,
\begin{equation}
    \Gamma^{WW}_{\rm lim} \simeq 1.5 \times 10^{19} s^{-1},
    \label{eq:ice-cube-romani}
\end{equation}
near the thermal Higgsino mass at $1.08$ TeV. We compare the resulting annihilation rate to the above limit as a representative check. We consider a model point to be excluded if the projected rate exceeds the limit of Eq.~(\ref{eq:ice-cube-romani}), log-interpolated in mass. We stress that this is a $W^+ W^-$-template constraint. Since the present model can also have other significant annihilation contribution, e.g., $\eta_2 \eta_2 \to hh$, applying the $WW(+ZZ)$ neutrino limit directly to the total annihilation rate would overestimate the constraint because the $hh$ channel produces a softer neutrino spectrum.

In the present model, $C_\odot \propto \sin^2\theta$, whereas the electroweak $WW/ZZ$ annihilation channels scale as $\langle\sigma v\rangle \propto \sin^4\theta$. Consequently, for the non-equilibrium regime, $t_\odot/\tau \ll 1$, the annihilation rate is $\Gamma_{\mathrm{ann}} \to \frac{t_\odot^2}{2} \frac{C_\odot^2\langle\sigma v\rangle_{\mathrm{tot}}}{V_{\mathrm{eff}}} \propto \sin^8\theta$. Therefore, the same small mixing angle that saturates the LZ rate further suppresses both solar capture and annihilation. This behavior is illustrated in Fig.~\ref{fig:solar-theta}, where we show $\Gamma_{\mathrm{ann}}/\Gamma_{\mathrm{lim}}^{WW}$ as a function of $\delta_A$ for the scan range $\theta \in [10^{-3}, 1]$, with the points color-coded according to $\theta$. The dashed magenta line corresponds to $\Gamma_{\mathrm{ann}}/\Gamma_{\mathrm{lim}}^{WW} = 1$, above which the corresponding point is excluded by the adopted IceCube limit. Here, larger $\theta$ corresponds to a larger doublet fraction, while smaller $\theta$ denotes increasingly singlet-like DM. The figure shows that the more singlet-like model points (i.e., $\theta \lesssim \mathcal{O}(0.05)$) generally yield a much smaller $\Gamma_{\mathrm{ann}}/\Gamma_{\mathrm{lim}}^{WW}$, whereas the more doublet-like model points approach or exceed the IceCube limit. All points shown satisfy the SI DD constraints from LZ-2025, reproduce the LZ recoil event, and remain $\Omega_{\rm DM} h^2 \leq 0.123$.

Fig. \ref{fig:ID} shows the same ratio $\Gamma_{\mathrm{ann}}/\Gamma_{\mathrm{lim}}^{WW}$ as a function of $\delta_A$, but for the narrower mixing-angle range $\theta \in [10^{-3}, 0.1]$. In addition, all model points satisfy $\pm 3\sigma$ relic-density constraint, including the grey points lying above the adopted IceCube limit. The blue and red points distinguish the non-thermal and fully thermalized treatments of the captured DM population, respectively. The non-thermal points generally lie well below the IceCube limit because the inefficient cooling leads to an extended captured population, increasing $V_{\mathrm{eff}}$ and suppressing annihilation. In contrast, forcing the population to thermalize reduces $V_{\mathrm{eff}}$, enhancing the annihilation rate and hence strengthening the solar constraint. This illustrates that the indirect constraint is controlled not only by the capture rate but also by the subsequent cooling and annihilation dynamics.

The dependence on $\delta_A$ follows from the interplay between endothermic kinematics and the LZ normalization. Increasing $\delta_A$ suppresses solar capture, but the inelastic LZ rate is simultaneously reduced, requiring a larger doublet fraction $\sin^2\theta$ to maintain $N_{\mathrm{SR}} \simeq 1$. Since the same singlet-doublet fraction controls the electroweak interactions relevant for solar capture and annihilation, the resulting solar rate reflects a competition between these effects.

The two benchmark points (see Table \ref{tab:benchmarks}) marked by stars lie safely below the adopted IceCube limit, with $\Gamma_{\mathrm{ann}}/\Gamma_{\mathrm{lim}} \simeq \mathcal{O}(10^{-2})$. These correspond to the navy-blue recoil spectra shown in Fig.~\ref{fig:2}(a), top and bottom panels, respectively. Both points are far from capture-annihilation equilibrium, with $t_{\odot}/\tau \simeq 0.09$, and consequently have $\tanh^2(t_{\odot}/\tau) \simeq 7 \times 10^{-3}$. Their suppressed annihilation rates are therefore reflected in their location well below the unity line in Fig.~\ref{fig:ID}.

\section{DISCUSSION AND CONCLUSIONS}
\label{sec4}
We have shown the possibility of explaining the recently reported high-energy recoil event by the LZ collaboration with singlet-doublet scalar DM. The required recoil is generated by inelastic up-scattering of DM into its heavier partner via $Z$-mediated process. The singlet-doublet mixing controls the relative proportion of singlet and doublet scalar components in the DM mass eigenstate thereby controlling the rates of processes responsible for DM relic and direct-detection. We find the parameter space consistent with observed DM relic and direct-detection bounds on elastic DM-nucleon scattering. We then evaluate the inelastic scattering cross-section and the corresponding event rate at LZ to fit the model parameter space with the recently reported event. We find a wide range of parameter space ranging from $\sim 400$ GeV to $\sim \mathcal{O}(1)$ TeV consistent with the LZ event for suitably adjusted mass splitting between DM and its heavier partner.

\begin{figure*}[!htpb]
    \centering
   \includegraphics[width=0.45\linewidth]{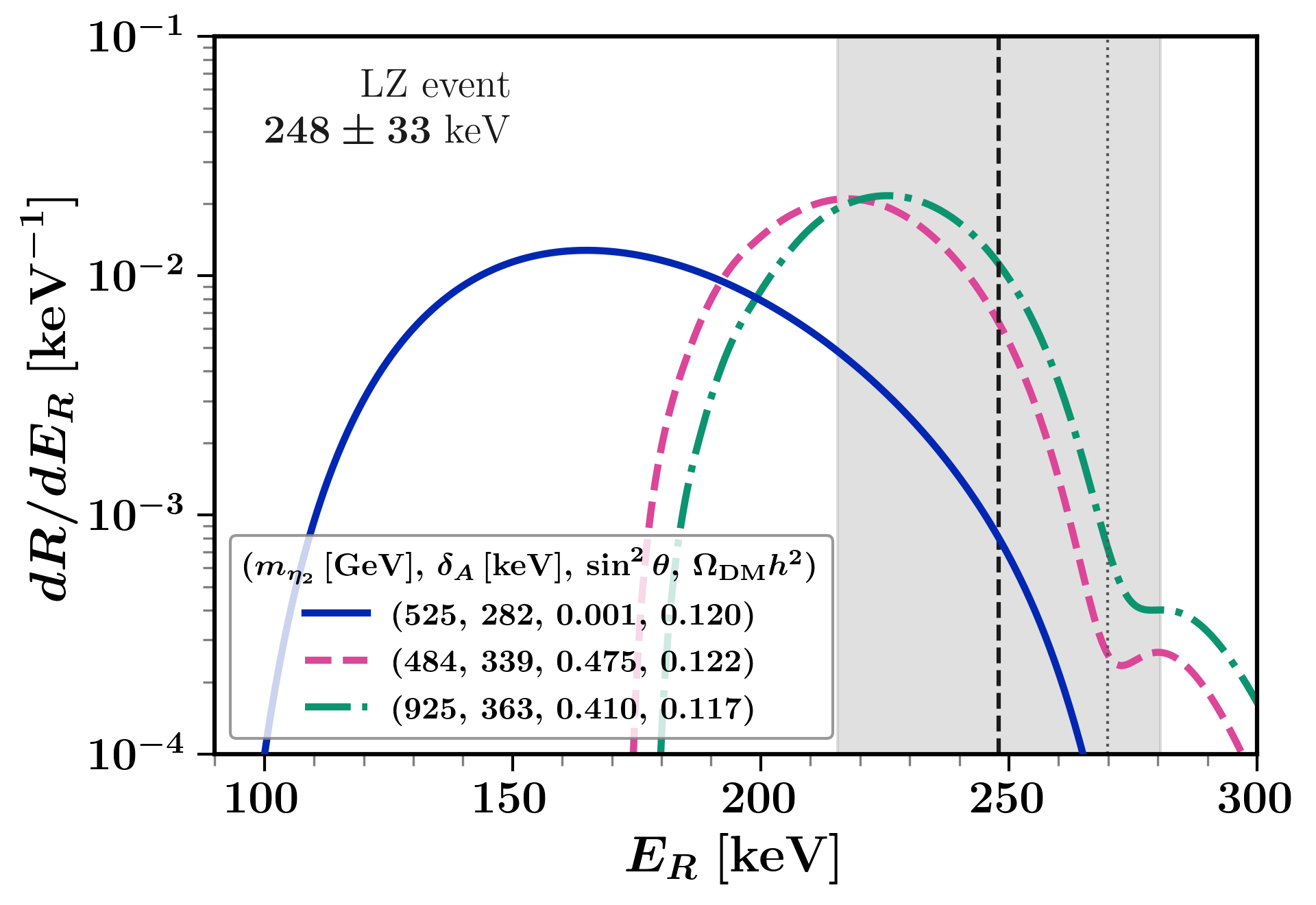}
    \hspace{0.7cm}
    \includegraphics[width=0.45\textwidth]{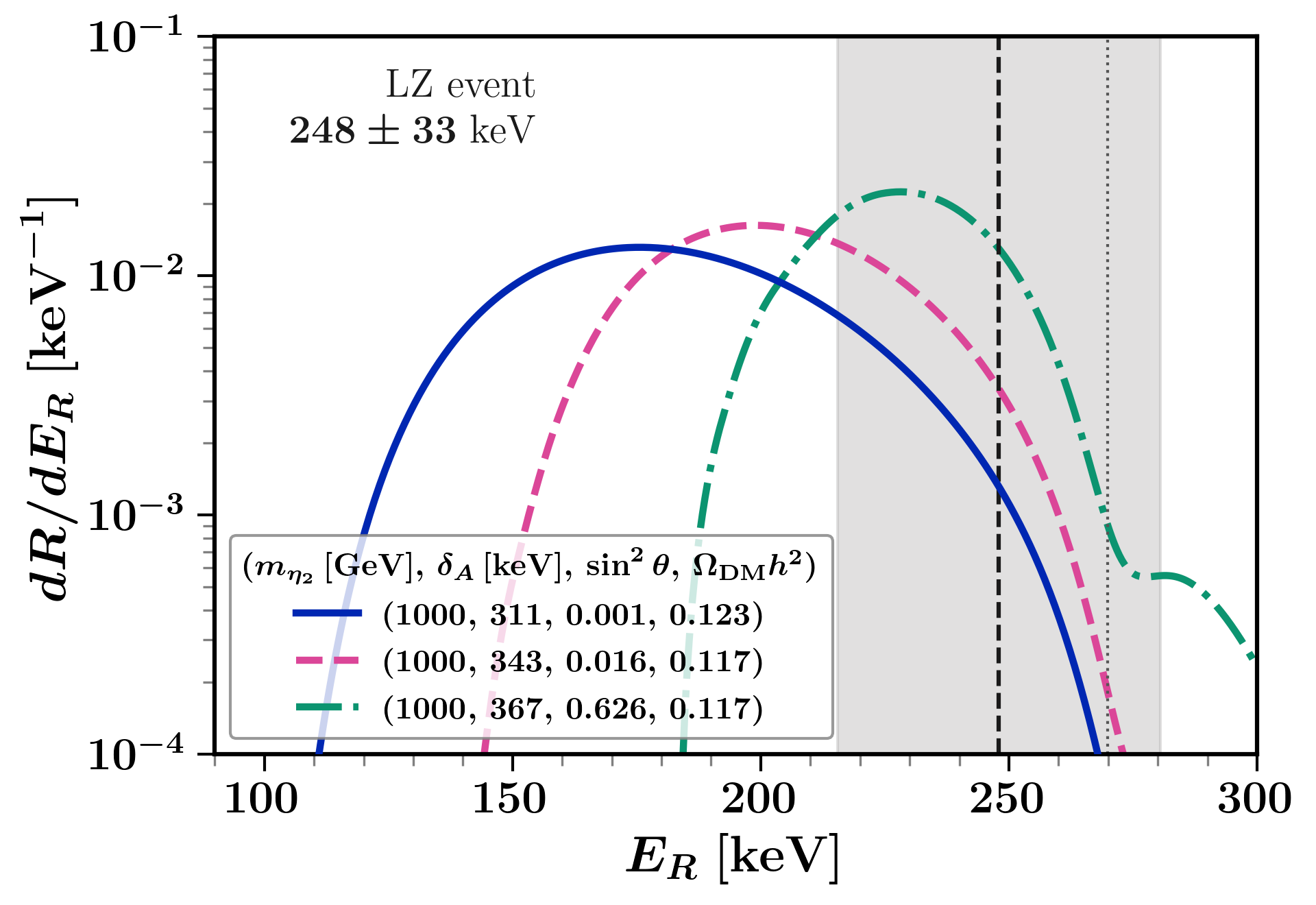}

    \vspace{-0.05cm}
    \makebox[0.48\textwidth][c]{(a)}
    \hfill
    \makebox[0.48\textwidth][c]{(b)}

    \caption{Differential recoil spectra, $dR/dE_R$, as a function of the recoil energy $E_R$ for the same benchmark points as in Fig.~2 without Gaussian energy smearing. The vertical dashed line denotes the central recoil energy, $E_R=248~\mathrm{keV}$, of the LZ event, while the grey shaded band indicates the corresponds to the interval $E_R=248\pm33~\mathrm{keV}$ \cite{LZ:2026axp}.}
    \label{fig:appx:true-recoil}
\end{figure*}

In this work, we have performed a detailed scan of the parameter space of our model framework and considered a $\chi^2$-based analysis requiring the number of signal event $N_{\rm SR}=1$. Across the DM mass range, it selects $\delta_A \in [271.6, 374.6] \text{ keV}$ and narrowing to $\delta_A \in [309.8, 374.8] \text{ keV}$ for $m_{\eta_2}=1\text{ TeV}$, with $60\%-80\%$ of the in-window rate falling within the signal region. The extended unbinned likelihood admits $0.1 \lesssim N_{\rm SR} \lesssim 3.6$ at 90\% CL. The singlet admixture supplies precisely this freedom: $\xi \sin^2\theta$ varies over three decades below the pure-doublet ceiling, allowing the model to populate the LZ two-sided $90\%$ CL band continuously across $\delta_A \simeq 310-375\text{ keV}$, whereas a pure inert doublet meets it only for $\delta_A \gtrsim 360\text{ keV}$, close to the kinematic wall $\delta_A \to \delta_{\rm max}$. The model thus accommodates the event over a far wider mass range than the $\gtrsim 550\text{ GeV}$ required of a pure doublet, while evading the elastic SI limits through the $\kappa$-induced blind spot in $\lambda_{\rm eff}$.

We further examined the complementary constraints from solar capture and subsequent DM annihilation into high-energy neutrinos. Despite the efficient capture enabled by the solar gravitational potential, the small doublet admixture that reproduces the LZ event strongly suppresses the annihilation signal. Only the more doublet-like points are tightly constrained from the solar-capture limits. For the considered benchmark points, the resulting annihilation rates remain below the representative IceCube limit, with $\Gamma_{\mathrm{ann}}/\Gamma_{\mathrm{lim}}^{WW} \sim \mathcal{O}(10^{-2})$.

While we stick to the minimal singlet-doublet scalar DM scenario in this work, the model can have several promising extensions, phenomenology and detection prospects. One possibility is to connect it to the origin of light neutrino mass by inclusion of $Z_2$-odd right-handed neutrinos (RHNs). This leads to the ScotoSinglet framework \cite{Beniwal:2020hjc} where light neutrino masses arise radiatively from the $Z_2$-odd sector in the scotogenic fashion \cite{Tao:1996vb, Ma:2006km}. The heavy RHNs can also generate the observed baryon asymmetry of the Universe via leptogenesis \cite{Hugle:2018qbw}. The combination of $Z_2$-odd singlet-doublet scalars can also play non-trivial role in leptogenesis via their Higgs portal or trilinear interactions \cite{Borah:2025hpo}. Additionally, the same $Z_2$-odd scalars can also assist in realizing a first-order electroweak phase transition \cite{Borah:2025hpo} which offers complementary detection prospects via stochastic gravitational waves. The sub-TeV $Z_2$-odd scalars can also lead to promising collider prospects \cite{Gustafsson:2012aj, Datta:2016nfz, Poulose:2016lvz, Miao:2010rg, Belyaev:2016lok, Belyaev:2018ext}. In addition, the compressed mass spectrum required for inelastic scattering of DM can have other interesting signatures like displaced vertex, charged tracks which can be probed at ongoing and future colliders \cite{Borah:2018smz, Blinov:2015qva}.

\acknowledgments
The work of D.B. is supported by the Science and Engineering Research Board (SERB), Government of India grant CRG/2022/000603. P. B. acknowledges the financial support received from the Indian Institute of Technology Guwahati (IITG) as an Institute Post-Doc Fellow (IPDF), grant {IITG/AR/IPDF/2026-27/024}.\\

\appendix
\section{True-recoil spectra and threshold kinematics}
\label{appx:true-recoil}


Fig.~\ref{fig:appx:true-recoil} shows the same benchmarks represented in Fig. \ref{fig:2} with the efficiency $\epsilon_{\rm LZ}(E_R)$ applied but without the Gaussian energy response, so that each spectrum terminates at the sharp endothermic edge $E_R^-$ below which the rate vanishes identically. The rise of $E_R^-$ with $\delta_A$, for both the benchmark groups in Fig.~\ref{fig:appx:true-recoil} (a) and (b), is what allows the $248\text{ keV}$ feature to select $\delta_A = \mathcal{O}(280-370)\text{ keV}$: the accessible interval $[E_R^-, E_R^+]$ must bracket the event. The structure near $270-290\text{ keV}$ reflects the Helm form-factor minimum for xenon; a full shell-model response fills this minimum and can shift the predicted rate in and above the ROI~\cite{Rodd:2026tyn, Su:2026rwz}, so the spectra there should be read as parametrization-dependent.


\begin{thebibliography}{10}

\bibitem{LZ:2026axp}
{\scshape LZ} collaboration, \emph{{Search for dark matter particle interactions in an extended nuclear recoil energy window with the LUX-ZEPLIN (LZ) experiment}},  \href{https://arxiv.org/abs/2609.02823}{{\ttfamily 2609.02823}}.

\bibitem{Jeesun:2026vzo}
S.~Jeesun and A.~Majumdar, \emph{{Atmospheric neutrino up-scattering explanation of LZ 2026 excess}},  \href{https://arxiv.org/abs/2609.04185}{{\ttfamily 2609.04185}}.

\bibitem{Fan:2026kxx}
J.~Fan and M.~Reece, \emph{{Higgsino Above the Sea of Fog}},  \href{https://arxiv.org/abs/2609.01504}{{\ttfamily 2609.01504}}.

\bibitem{Wu:2026nhi}
L.~Wu, Y.~Zhang and B.~Zhu, \emph{{TeV Higgsino Dark Matter from LZ Nuclear Recoil to Fermi-LAT Gamma Rays}},  \href{https://arxiv.org/abs/2609.01590}{{\ttfamily 2609.01590}}.

\bibitem{Freese:2026sga}
K.~Freese and D.P.~Theodosopoulos, \emph{{Higgsino Dark Matter Interpretation of the LUX-ZEPLIN 248 keV Nuclear-Recoil Event}},  \href{https://arxiv.org/abs/2609.01583}{{\ttfamily 2609.01583}}.

\bibitem{Du:2026guj}
X.~Du and F.~Wang, \emph{{TeV Higgsino Interpretation of the LZ High-Recoil Event with Intermediate-Scale Electroweak Gauginos}},  \href{https://arxiv.org/abs/2609.04163}{{\ttfamily 2609.04163}}.

\bibitem{Yin:2026jnn}
W.~Yin, \emph{{A PQ-Symmetric High-Scale SUSY Interpretation of the LZ High-Energy Recoil}},  \href{https://arxiv.org/abs/2609.01892}{{\ttfamily 2609.01892}}.

\bibitem{McCabe:2026crm}
C.~McCabe, \emph{{Seasonal dark matter from the LUX-ZEPLIN high-energy event}},  \href{https://arxiv.org/abs/2609.04181}{{\ttfamily 2609.04181}}.

\bibitem{Unwin:2026rdp}
J.~Unwin, \emph{{Axion Portal Dark Matter and the LUX-ZEPLIN High-Recoil Event}},  \href{https://arxiv.org/abs/2609.04186}{{\ttfamily 2609.04186}}.

\bibitem{Smirnov:2026aqk}
J.~Smirnov, S.~Griffith and J.F.~Beacom, \emph{{Inelastic Signatures of Electroweak Dark Matter}},  \href{https://arxiv.org/abs/2609.04144}{{\ttfamily 2609.04144}}.

\bibitem{Nomura:2026qyq}
Y.~Nomura, \emph{{Dark Matter as the $Z_2$ Partner of the Standard Model Higgs Boson}},  \href{https://arxiv.org/abs/2609.02505}{{\ttfamily 2609.02505}}.

\bibitem{Lou:2026idn}
Y.~Lou and C.-T.~Lu, \emph{{Fermionic Dark Matter Absorption and the High-Energy Event in LUX-ZEPLIN}},  \href{https://arxiv.org/abs/2609.01592}{{\ttfamily 2609.01592}}.

\bibitem{Su:2026rwz}
L.~Su, J.M.~Yang and W.-N.~Yang, \emph{{Inelastic Dark Matter Signature at High Recoil Energy in LUX-ZEPLIN and CRESST}},  \href{https://arxiv.org/abs/2609.01475}{{\ttfamily 2609.01475}}.

\bibitem{DiMauro:2026ldr}
M.~Di~Mauro, \emph{{Dark Matter at the Kinematic Edge: Interpreting the 248 keV LZ Nuclear-Recoil Candidate}},  \href{https://arxiv.org/abs/2609.02608}{{\ttfamily 2609.02608}}.

\bibitem{Yamashita:2026ump}
K.~Yamashita, \emph{{Inelastic Dark Photon Dark Matter for the LUX-ZEPLIN High-Recoil Event and the Galactic Halo Gamma-Ray Excess}},  \href{https://arxiv.org/abs/2609.02868}{{\ttfamily 2609.02868}}.

\bibitem{Chattopadhyay:2026ryw}
U.~Chattopadhyay, D.~Das, R.~Puri and J.~Roy, \emph{{Sub-TeV Singlino Dark Matter in light from Sagittarius A$^\ast$ and LUX-ZEPLIN Nuclear-Recoil Event}},  \href{https://arxiv.org/abs/2609.02994}{{\ttfamily 2609.02994}}.

\bibitem{deLima:2026shq}
C.H.~de~Lima, \emph{{Exothermic Dark Matter at LZ}},  \href{https://arxiv.org/abs/2609.05204}{{\ttfamily 2609.05204}}.

\bibitem{Visinelli:2026kgt}
L.~Visinelli, \emph{{A Peccei-Quinn Origin for Inelastic Electroweak Dark Matter after LUX-ZEPLIN}},  \href{https://arxiv.org/abs/2609.02807}{{\ttfamily 2609.02807}}.

\bibitem{Wang:2026ytg}
L.~Wang and Y.~Xiao, \emph{{The Inert Doublet Model of Dark Matter and the LUX-ZEPLIN High-Recoil Event}},  \href{https://arxiv.org/abs/2609.06571}{{\ttfamily 2609.06571}}.

\bibitem{Dent:2026bji}
J.B.~Dent and J.L.~Newstead, \emph{{Exothermic and Endothermic Inelastic Dark Matter Interpretations at LZ: Sideband Constraints and Future Prospects}},  \href{https://arxiv.org/abs/2609.04673}{{\ttfamily 2609.04673}}.

\bibitem{Gu:2026vto}
G.~Gu, L.~Li, S.-S.~Tang and Y.~Xu, \emph{{Inelastic from the Other Side: Xenon Excitation Signals in Light of the LZ High-Recoil Event}},  \href{https://arxiv.org/abs/2609.05291}{{\ttfamily 2609.05291}}.

\bibitem{Pospelov:2026ewn}
M.~Pospelov and H.~Ramani, \emph{{Strong Constraints on Higgsino Dark Matter from Solar Capture}},  \href{https://arxiv.org/abs/2609.02775}{{\ttfamily 2609.02775}}.

\bibitem{Bose:2026ndd}
D.~Bose et~al., \emph{{Not so good $\nu$s for Higgsino dark matter as LZ excess: stringent limits from Super-Kamiokande and IceCube}},  \href{https://arxiv.org/abs/2609.07807}{{\ttfamily 2609.07807}}.

\bibitem{Nguyen:2026lui}
T.T.Q.~Nguyen, T.~Linden and D.~Hooper, \emph{{Solar Neutrino Constraints on Inelastic Dark Matter Scattering in Light of Recent LUX-ZEPLIN Observations}},  \href{https://arxiv.org/abs/2609.11833}{{\ttfamily 2609.11833}}.

\bibitem{DiMauro:2026dqp}
M.~Di~Mauro and H.~Shaikh, \emph{{Solar Capture Tests of Inelastic Dark Matter after the LZ High-Recoil Event}},  \href{https://arxiv.org/abs/2609.06760}{{\ttfamily 2609.06760}}.

\bibitem{Rodd:2026tyn}
N.L.~Rodd, B.R.~Safdi, T.R.~Slatyer and W.L.~Xu, \emph{{Confronting the Higgsino Interpretation of the LZ Event with the High-Energy Sideband}},  \href{https://arxiv.org/abs/2609.04175}{{\ttfamily 2609.04175}}.

\bibitem{Beniwal:2020hjc}
A.~Beniwal, J.~Herrero-Garcia, N.~Leerdam, M.~White and A.G.~Williams, \emph{{The ScotoSinglet Model: a scalar singlet extension of the Scotogenic Model}}, \href{https://doi.org/10.1007/JHEP06(2021)136}{\emph{JHEP} {\bfseries 21} (2020) 136} [\href{https://arxiv.org/abs/2010.05937}{{\ttfamily 2010.05937}}].

\bibitem{Ma:2006km}
E.~Ma, \emph{{Verifiable radiative seesaw mechanism of neutrino mass and dark matter}}, \href{https://doi.org/10.1103/PhysRevD.73.077301}{\emph{Phys. Rev. D} {\bfseries 73} (2006) 077301} [\href{https://arxiv.org/abs/hep-ph/0601225}{{\ttfamily hep-ph/0601225}}].

\bibitem{Cirelli:2005uq}
M.~Cirelli, N.~Fornengo and A.~Strumia, \emph{{Minimal dark matter}}, \href{https://doi.org/10.1016/j.nuclphysb.2006.07.012}{\emph{Nucl. Phys. B} {\bfseries 753} (2006) 178} [\href{https://arxiv.org/abs/hep-ph/0512090}{{\ttfamily hep-ph/0512090}}].

\bibitem{Barbieri:2006dq}
R.~Barbieri, L.J.~Hall and V.S.~Rychkov, \emph{{Improved naturalness with a heavy Higgs: An Alternative road to LHC physics}}, \href{https://doi.org/10.1103/PhysRevD.74.015007}{\emph{Phys. Rev. D} {\bfseries 74} (2006) 015007} [\href{https://arxiv.org/abs/hep-ph/0603188}{{\ttfamily hep-ph/0603188}}].

\bibitem{LopezHonorez:2006gr}
L.~Lopez~Honorez, E.~Nezri, J.F.~Oliver and M.H.G.~Tytgat, \emph{{The Inert Doublet Model: An Archetype for Dark Matter}}, \href{https://doi.org/10.1088/1475-7516/2007/02/028}{\emph{JCAP} {\bfseries 02} (2007) 028} [\href{https://arxiv.org/abs/hep-ph/0612275}{{\ttfamily hep-ph/0612275}}].

\bibitem{Hambye:2009pw}
T.~Hambye, F.S.~Ling, L.~Lopez~Honorez and J.~Rocher, \emph{{Scalar Multiplet Dark Matter}}, \href{https://doi.org/10.1007/JHEP05(2010)066}{\emph{JHEP} {\bfseries 07} (2009) 090} [\href{https://arxiv.org/abs/0903.4010}{{\ttfamily 0903.4010}}].

\bibitem{Dolle:2009fn}
E.M.~Dolle and S.~Su, \emph{{The Inert Dark Matter}}, \href{https://doi.org/10.1103/PhysRevD.80.055012}{\emph{Phys. Rev. D} {\bfseries 80} (2009) 055012} [\href{https://arxiv.org/abs/0906.1609}{{\ttfamily 0906.1609}}].

\bibitem{LopezHonorez:2010eeh}
L.~Lopez~Honorez and C.E.~Yaguna, \emph{{The inert doublet model of dark matter revisited}}, \href{https://doi.org/10.1007/JHEP09(2010)046}{\emph{JHEP} {\bfseries 09} (2010) 046} [\href{https://arxiv.org/abs/1003.3125}{{\ttfamily 1003.3125}}].

\bibitem{LopezHonorez:2010tb}
L.~Lopez~Honorez and C.E.~Yaguna, \emph{{A new viable region of the inert doublet model}}, \href{https://doi.org/10.1088/1475-7516/2011/01/002}{\emph{JCAP} {\bfseries 01} (2011) 002} [\href{https://arxiv.org/abs/1011.1411}{{\ttfamily 1011.1411}}].

\bibitem{Borah:2012pu}
D.~Borah and J.M.~Cline, \emph{{Inert Doublet Dark Matter with Strong Electroweak Phase Transition}}, \href{https://doi.org/10.1103/PhysRevD.86.055001}{\emph{Phys. Rev. D} {\bfseries 86} (2012) 055001} [\href{https://arxiv.org/abs/1204.4722}{{\ttfamily 1204.4722}}].

\bibitem{Dasgupta:2014hha}
A.~Dasgupta and D.~Borah, \emph{{Scalar Dark Matter with Type II Seesaw}}, \href{https://doi.org/10.1016/j.nuclphysb.2014.10.028}{\emph{Nucl. Phys. B} {\bfseries 889} (2014) 637} [\href{https://arxiv.org/abs/1404.5261}{{\ttfamily 1404.5261}}].

\bibitem{Borah:2017dfn}
D.~Borah and A.~Gupta, \emph{{New viable region of an inert Higgs doublet dark matter model with scotogenic extension}}, \href{https://doi.org/10.1103/PhysRevD.96.115012}{\emph{Phys. Rev. D} {\bfseries 96} (2017) 115012} [\href{https://arxiv.org/abs/1706.05034}{{\ttfamily 1706.05034}}].

\bibitem{Silveira:1985rk}
V.~Silveira and A.~Zee, \emph{{SCALAR PHANTOMS}}, \href{https://doi.org/10.1016/0370-2693(85)90624-0}{\emph{Phys. Lett. B} {\bfseries 161} (1985) 136}.

\bibitem{McDonald:1993ex}
J.~McDonald, \emph{{Gauge singlet scalars as cold dark matter}}, \href{https://doi.org/10.1103/PhysRevD.50.3637}{\emph{Phys. Rev. D} {\bfseries 50} (1994) 3637} [\href{https://arxiv.org/abs/hep-ph/0702143}{{\ttfamily hep-ph/0702143}}].

\bibitem{Belanger:2026asz}
G.~Belanger, A.~Belyaev, N.~Bernal, F.~Boudjema, S.~Chakraborti, A.~Goudelis et~al., \emph{{micrOMEGAs 7: Beyond standard cosmology}},  \href{https://arxiv.org/abs/2606.06645}{{\ttfamily 2606.06645}}.

\bibitem{Alguero:2023zol}
G.~Alguero, G.~Belanger, F.~Boudjema, S.~Chakraborti, A.~Goudelis, S.~Kraml et~al., \emph{{micrOMEGAs 6.0: N-component dark matter}}, \href{https://doi.org/10.1016/j.cpc.2024.109133}{\emph{Comput. Phys. Commun.} {\bfseries 299} (2024) 109133} [\href{https://arxiv.org/abs/2312.14894}{{\ttfamily 2312.14894}}].

\bibitem{Belanger:2018ccd}
G.~Belanger, F.~Boudjema, A.~Goudelis, A.~Pukhov and B.~Zaldivar, \emph{{micrOMEGAs5.0 : Freeze-in}}, \href{https://doi.org/10.1016/j.cpc.2018.04.027}{\emph{Comput. Phys. Commun.} {\bfseries 231} (2018) 173} [\href{https://arxiv.org/abs/1801.03509}{{\ttfamily 1801.03509}}].

\bibitem{Alloul:2013bka}
A.~Alloul, N.D.~Christensen, C.~Degrande, C.~Duhr and B.~Fuks, \emph{{FeynRules 2.0 - A complete toolbox for tree-level phenomenology}}, \href{https://doi.org/10.1016/j.cpc.2014.04.012}{\emph{Comput. Phys. Commun.} {\bfseries 185} (2014) 2250} [\href{https://arxiv.org/abs/1310.1921}{{\ttfamily 1310.1921}}].

\bibitem{Bechtle:2020pkv}
P.~Bechtle, D.~Dercks, S.~Heinemeyer, T.~Klingl, T.~Stefaniak, G.~Weiglein et~al., \emph{{HiggsBounds-5: Testing Higgs Sectors in the LHC 13 TeV Era}}, \href{https://doi.org/10.1140/epjc/s10052-020-08557-9}{\emph{Eur. Phys. J. C} {\bfseries 80} (2020) 1211} [\href{https://arxiv.org/abs/2006.06007}{{\ttfamily 2006.06007}}].

\bibitem{Bechtle:2020uwn}
P.~Bechtle, S.~Heinemeyer, T.~Klingl, T.~Stefaniak, G.~Weiglein and J.~Wittbrodt, \emph{{HiggsSignals-2: Probing new physics with precision Higgs measurements in the LHC 13 TeV era}}, \href{https://doi.org/10.1140/epjc/s10052-021-08942-y}{\emph{Eur. Phys. J. C} {\bfseries 81} (2021) 145} [\href{https://arxiv.org/abs/2012.09197}{{\ttfamily 2012.09197}}].

\bibitem{LZ:2024zvo}
{\scshape LZ} collaboration, \emph{{Dark Matter Search Results from 4.2{\,}{\,}Tonne-Years of Exposure of the LUX-ZEPLIN (LZ) Experiment}}, \href{https://doi.org/10.1103/4dyc-z8zf}{\emph{Phys. Rev. Lett.} {\bfseries 135} (2025) 011802} [\href{https://arxiv.org/abs/2410.17036}{{\ttfamily 2410.17036}}].

\bibitem{DARWIN:2016hyl}
{\scshape DARWIN} collaboration, \emph{{DARWIN: towards the ultimate dark matter detector}}, \href{https://doi.org/10.1088/1475-7516/2016/11/017}{\emph{JCAP} {\bfseries 11} (2016) 017} [\href{https://arxiv.org/abs/1606.07001}{{\ttfamily 1606.07001}}].

\bibitem{Planck:2018vyg}
{\scshape Planck} collaboration, \emph{{Planck 2018 results. VI. Cosmological parameters}}, \href{https://doi.org/10.1051/0004-6361/201833910}{\emph{Astron. Astrophys.} {\bfseries 641} (2020) A6} [\href{https://arxiv.org/abs/1807.06209}{{\ttfamily 1807.06209}}].

\bibitem{Smith_2001}
D.~Smith and N.~Weiner, \emph{Inelastic dark matter}, \href{https://doi.org/10.1103/physrevd.64.043502}{\emph{Physical Review D} {\bfseries 64} (2001) }.

\bibitem{Tucker_Smith_2005}
D.~Tucker-Smith and N.~Weiner, \emph{Status of inelastic dark matter}, \href{https://doi.org/10.1103/physrevd.72.063509}{\emph{Physical Review D} {\bfseries 72} (2005) }.

\bibitem{Baxter_2021}
D.~Baxter, I.M.~Bloch, E.~Bodnia, X.~Chen, J.~Conrad, P.~Di~Gangi et~al., \emph{Recommended conventions for reporting results from direct dark matter searches}, \href{https://doi.org/10.1140/epjc/s10052-021-09655-y}{\emph{The European Physical Journal C} {\bfseries 81} (2021) }.

\bibitem{Smith:2006ym}
M.C.~Smith et~al., \emph{{The RAVE Survey: Constraining the Local Galactic Escape Speed}}, \href{https://doi.org/10.1111/j.1365-2966.2007.11964.x}{\emph{Mon. Not. Roy. Astron. Soc.} {\bfseries 379} (2007) 755} [\href{https://arxiv.org/abs/astro-ph/0611671}{{\ttfamily astro-ph/0611671}}].

\bibitem{LEWIN199687}
J.~Lewin and P.~Smith, \emph{Review of mathematics, numerical factors, and corrections for dark matter experiments based on elastic nuclear recoil}, \href{https://doi.org/https://doi.org/10.1016/S0927-6505(96)00047-3}{\emph{Astroparticle Physics} {\bfseries 6} (1996) 87}.

\bibitem{Lewin:1995rx}
J.D.~Lewin and P.F.~Smith, \emph{{Review of mathematics, numerical factors, and corrections for dark matter experiments based on elastic nuclear recoil}}, \href{https://doi.org/10.1016/S0927-6505(96)00047-3}{\emph{Astropart. Phys.} {\bfseries 6} (1996) 87}.

\bibitem{WIMpy-code}
B.J.~Kavanagh and T.D.P.~Edwards, \emph{\textnormal{WIMpy\_NREFT v1.2 [Computer Software]}, \href{https://doi.org/10.5281/zenodo.1230503}{\textnormal{doi:10.5281/zenodo.1230503}}\textnormal{. Available at }\url{https://github.com/bradkav/WIMpy_NREFT}},  2024.

\bibitem{Jeong:2021bpl}
I.~Jeong, S.~Kang, S.~Scopel and G.~Tomar, \emph{{WimPyDD: An object{\textendash}oriented Python code for the calculation of WIMP direct detection signals}}, \href{https://doi.org/10.1016/j.cpc.2022.108342}{\emph{Comput. Phys. Commun.} {\bfseries 276} (2022) 108342} [\href{https://arxiv.org/abs/2106.06207}{{\ttfamily 2106.06207}}].

\bibitem{IceCube:2025fcu}
{\scshape IceCube} collaboration, \emph{{Search for High-Energy Neutrinos From the Sun Using Ten Years of IceCube Data}},  \href{https://arxiv.org/abs/2507.08457}{{\ttfamily 2507.08457}}.

\bibitem{Press:1985ug}
W.H.~Press and D.N.~Spergel, \emph{{Capture by the sun of a galactic population of weakly interacting massive particles}}, \href{https://doi.org/10.1086/163485}{\emph{Astrophys. J.} {\bfseries 296} (1985) 679}.

\bibitem{Gould:1987ir}
A.~Gould, \emph{{Resonant Enhancements in WIMP Capture by the Earth}}, \href{https://doi.org/10.1086/165653}{\emph{Astrophys. J.} {\bfseries 321} (1987) 571}.

\bibitem{Bahcall:2004pz}
J.N.~Bahcall, A.M.~Serenelli and S.~Basu, \emph{{New solar opacities, abundances, helioseismology, and neutrino fluxes}}, \href{https://doi.org/10.1086/428929}{\emph{Astrophys. J. Lett.} {\bfseries 621} (2005) L85} [\href{https://arxiv.org/abs/astro-ph/0412440}{{\ttfamily astro-ph/0412440}}].

\bibitem{Lee:2026jxl}
S.J.~Lee and T.~Youn, \emph{{Mixing-suppressed inelastic dark matter: a minimal model for the LZ 248 keV event}},  \href{https://arxiv.org/abs/2609.09138}{{\ttfamily 2609.09138}}.

\bibitem{Tao:1996vb}
Z.-j.~Tao, \emph{{Radiative seesaw mechanism at weak scale}}, \href{https://doi.org/10.1103/PhysRevD.54.5693}{\emph{Phys. Rev. D} {\bfseries 54} (1996) 5693} [\href{https://arxiv.org/abs/hep-ph/9603309}{{\ttfamily hep-ph/9603309}}].

\bibitem{Hugle:2018qbw}
T.~Hugle, M.~Platscher and K.~Schmitz, \emph{{Low-Scale Leptogenesis in the Scotogenic Neutrino Mass Model}}, \href{https://doi.org/10.1103/PhysRevD.98.023020}{\emph{Phys. Rev. D} {\bfseries 98} (2018) 023020} [\href{https://arxiv.org/abs/1804.09660}{{\ttfamily 1804.09660}}].

\bibitem{Borah:2025hpo}
D.~Borah, D.~Mahanta and I.~Saha, \emph{{Gravitational wave signatures of dark sector portal leptogenesis}},  \href{https://arxiv.org/abs/2504.14671}{{\ttfamily 2504.14671}}.

\bibitem{Gustafsson:2012aj}
M.~Gustafsson, S.~Rydbeck, L.~Lopez-Honorez and E.~Lundstrom, \emph{{Status of the Inert Doublet Model and the Role of multileptons at the LHC}}, \href{https://doi.org/10.1103/PhysRevD.86.075019}{\emph{Phys. Rev. D} {\bfseries 86} (2012) 075019} [\href{https://arxiv.org/abs/1206.6316}{{\ttfamily 1206.6316}}].

\bibitem{Datta:2016nfz}
A.~Datta, N.~Ganguly, N.~Khan and S.~Rakshit, \emph{{Exploring collider signatures of the inert Higgs doublet model}}, \href{https://doi.org/10.1103/PhysRevD.95.015017}{\emph{Phys. Rev. D} {\bfseries 95} (2017) 015017} [\href{https://arxiv.org/abs/1610.00648}{{\ttfamily 1610.00648}}].

\bibitem{Poulose:2016lvz}
P.~Poulose, S.~Sahoo and K.~Sridhar, \emph{{Exploring the Inert Doublet Model through the dijet plus missing transverse energy channel at the LHC}}, \href{https://doi.org/10.1016/j.physletb.2016.12.022}{\emph{Phys. Lett. B} {\bfseries 765} (2017) 300} [\href{https://arxiv.org/abs/1604.03045}{{\ttfamily 1604.03045}}].

\bibitem{Miao:2010rg}
X.~Miao, S.~Su and B.~Thomas, \emph{{Trilepton Signals in the Inert Doublet Model}}, \href{https://doi.org/10.1103/PhysRevD.82.035009}{\emph{Phys. Rev. D} {\bfseries 82} (2010) 035009} [\href{https://arxiv.org/abs/1005.0090}{{\ttfamily 1005.0090}}].

\bibitem{Belyaev:2016lok}
A.~Belyaev, G.~Cacciapaglia, I.P.~Ivanov, F.~Rojas-Abatte and M.~Thomas, \emph{{Anatomy of the Inert Two Higgs Doublet Model in the light of the LHC and non-LHC Dark Matter Searches}}, \href{https://doi.org/10.1103/PhysRevD.97.035011}{\emph{Phys. Rev. D} {\bfseries 97} (2018) 035011} [\href{https://arxiv.org/abs/1612.00511}{{\ttfamily 1612.00511}}].

\bibitem{Belyaev:2018ext}
A.~Belyaev, T.R.~Fernandez Perez~Tomei, P.G.~Mercadante, C.S.~Moon, S.~Moretti, S.F.~Novaes et~al., \emph{{Advancing LHC probes of dark matter from the inert two-Higgs-doublet model with the monojet signal}}, \href{https://doi.org/10.1103/PhysRevD.99.015011}{\emph{Phys. Rev. D} {\bfseries 99} (2019) 015011} [\href{https://arxiv.org/abs/1809.00933}{{\ttfamily 1809.00933}}].

\bibitem{Borah:2018smz}
D.~Borah, D.~Nanda, N.~Narendra and N.~Sahu, \emph{{Right-handed neutrino dark matter with radiative neutrino mass in gauged B-L model}}, \href{https://doi.org/10.1016/j.nuclphysb.2019.114841}{\emph{Nucl. Phys. B} {\bfseries 950} (2020) 114841} [\href{https://arxiv.org/abs/1810.12920}{{\ttfamily 1810.12920}}].

\bibitem{Blinov:2015qva}
N.~Blinov, J.~Kozaczuk, D.E.~Morrissey and A.~de~la Puente, \emph{{Compressing the Inert Doublet Model}}, \href{https://doi.org/10.1103/PhysRevD.93.035020}{\emph{Phys. Rev. D} {\bfseries 93} (2016) 035020} [\href{https://arxiv.org/abs/1510.08069}{{\ttfamily 1510.08069}}].

\end{thebibliography}

\providecommand{\href}[2]{#2}\begingroup\raggedright\endgroup

\end{document}